\documentclass[acmsmall,nonacm,natbib=false]{acmart}

\usepackage{amsmath,amsfonts}
\usepackage{algorithmic}
\usepackage{graphicx}
\usepackage{textcomp}
\usepackage[dvipsnames]{xcolor}
\usepackage{subcaption}
\usepackage{booktabs}
\usepackage{multirow}
\usepackage{xurl}

\usepackage{pgfplots}
\pgfplotsset{compat=1.18}
\usepgfplotslibrary{groupplots}
\usetikzlibrary{arrows.meta,positioning,shapes.geometric,fit,backgrounds,calc}

\newcommand{\plotfont}{\sffamily}
\tikzset{every picture/.append style={font=\plotfont\small}}

\pgfplotsset{
  every axis/.append style={
    width=\columnwidth, height=0.62\columnwidth,
    grid=both, grid style={gray!25},
    tick label style={/pgf/number format/1000 sep={}},
    legend style={at={(0.03,0.97)}, anchor=north west,
                  draw=gray!60, fill=white, cells={anchor=west}},
  },
}

\definecolor{amdred}{RGB}{237,28,36}
\definecolor{nvgreen}{RGB}{118,185,0}

\pgfplotsset{
  discard if not/.style 2 args={
    x filter/.code={%
      \edef\tempa{\thisrow{#1}}%
      \edef\tempb{#2}%
      \ifx\tempa\tempb\else\fi
    }
  }
}

\pgfplotsset{
  annotated series/.style={
    thick, mark size=2pt,
    point meta=explicit symbolic,
    nodes near coords={\pgfplotspointmeta},
    visualization depends on={value \thisrow{anchor}\as\labelanchor},
    visualization depends on={value \thisrow{xshift}\as\labelxshift},
    visualization depends on={value \thisrow{yshift}\as\labelyshift},
    every node near coord/.append style={
      anchor=\labelanchor,
      xshift=\labelxshift pt,
      yshift=\labelyshift pt,
    },
  },
}

\definecolor{baselinecol}{RGB}{31,119,180}
\definecolor{solutioncol}{RGB}{214,39,40}
\definecolor{ttftpfifty}{HTML}{253494}
\definecolor{ttftpninety}{HTML}{2C7FB8}
\definecolor{ttftpninetynine}{HTML}{41B6C4}

\definecolor{kbGemm}{HTML}{1F77B4}        
\definecolor{kbRccl}{HTML}{FF7F0E}        
\definecolor{kbAttn}{HTML}{2CA02C}        
\definecolor{kbGather}{HTML}{D62728}      
\definecolor{kbActMul}{HTML}{9467BD}      
\definecolor{kbMoeSort}{HTML}{8C564B}     
\definecolor{kbQuant}{HTML}{E377C2}       
\definecolor{kbOther}{HTML}{7F7F7F}       
\definecolor{kbLayernorm}{HTML}{BCBD22}   
\definecolor{kbElementwise}{HTML}{17BECF} 
\definecolor{kbRotary}{HTML}{C5B0D5}      
\definecolor{kbReshape}{HTML}{C49C94}     
\definecolor{kbPagedAttn}{HTML}{FFBB78}   
\definecolor{kbPagedRed}{HTML}{F7B6D2}    

\pgfplotsset{
  kb/gemm/.style={fill=kbGemm, draw=kbGemm},
  kb/rccl/.style={fill=kbRccl, draw=kbRccl},
  kb/fwd_attn/.style={fill=kbAttn, draw=kbAttn},
  kb/mqa_attn/.style={fill=kbAttn, draw=kbAttn},
  kb/gather/.style={fill=kbGather, draw=kbGather},
  kb/act_and_mul/.style={fill=kbActMul, draw=kbActMul},
  kb/moe_sort/.style={fill=kbMoeSort, draw=kbMoeSort},
  kb/elementwise/.style={fill=kbElementwise, draw=kbElementwise},
  kb/layernorm/.style={fill=kbLayernorm, draw=kbLayernorm},
  kb/scaled_fp8_quant/.style={fill=kbQuant, draw=kbQuant},
  kb/rotary/.style={fill=kbRotary, draw=kbRotary},
  kb/reshape/.style={fill=kbReshape, draw=kbReshape},
  kb/paged_attn/.style={fill=kbPagedAttn, draw=kbPagedAttn},
  kb/paged_attn_reduce/.style={fill=kbPagedRed, draw=kbPagedRed},
  kb/other/.style={fill=kbOther, draw=kbOther},
}

\pgfplotsset{
  kernel stack panel/.style={
    width=\linewidth, height=1.15\linewidth,
    ybar stacked, bar width=5pt,
    xlabel={Input-length bucket},
    xtick=data, enlarge x limits=0.12,
    xmajorgrids=false,
    xticklabel style={rotate=45, anchor=north east, inner sep=1pt,
                      font=\plotfont\tiny},
    yticklabel style={font=\plotfont\scriptsize},
    ylabel style={font=\plotfont\scriptsize},
    xlabel style={font=\plotfont\scriptsize},
    ymin=0, ymax=100, ytick={0,20,40,60,80,100},
    legend style={at={(0.5,1.02)}, anchor=south, draw=none, fill=none,
                  column sep=0.3em, inner sep=1pt, font=\plotfont\tiny},
    legend cell align=left,
    legend image code/.code={\draw[##1] (0cm,-0.06cm) rectangle (0.16cm,0.06cm);},
  },
}

\pgfplotsset{
  qps bar panel/.style={
    width=\linewidth, height=0.68\linewidth,
    xlabel={Request rate (QPS)},
    ybar,
    symbolic x coords={0.1,0.2,0.3}, xtick={0.1,0.2,0.3},
    xmajorgrids=false,
    legend style={at={(0.5,1.02)}, anchor=south, draw=none, fill=none,
                  column sep=0.6em, font=\sffamily\scriptsize},
  },
}

\RequirePackage[
  datamodel=acmdatamodel,
  maxnames=200,
  style=acmnumeric
]{biblatex}
\newcommand{\takeawaybox}[1]{%
    \vspace{3pt}
    \noindent
    \fcolorbox{black!75}{gray!10}{%
        \parbox{\dimexpr\columnwidth-2\fboxsep-2\fboxrule}{%
            \setlength{\labelsep}{3pt}%
            \setlength{\leftmargini}{9pt}%
            \setlength{\listisep}{4pt}%
            \setlength{\partopsep}{0pt}%
            \setlength{\itemsep}{0pt}%
            \setlength{\parsep}{0pt}%
            \textbf{Lessons Learned:} \small #1
        }%
    }%
    \vspace{0pt}
}

\begin{document}


\def\CommentAP#1{\textcolor{red}{#1}}
\def\Edit#1{{#1}}
\def\Highlight#1{{#1}}
\def\Comment#1{\textcolor{MidnightBlue}{#1}}
\def\TODO#1{\textcolor{Red}{#1}}
\newcommand{\llamainf}{LLaMA3-8B inference}
\newcommand{\sdxl}{Stable Diffusion}
\newcommand{\qwen}{Qwen3-14B inference}
\newcommand{\llamatrain}{LLaMA2-3B training}
\newcommand{\qwenmodel}{Qwen3-14B}
\newcommand{\llamamodel}{LLaMA3-8B}
\newcommand{\mixtralmodel}{Mixtral-8x7B}
\newcommand{\twoten}{AMD Instinct\texttrademark~MI210}
\newcommand{\threeA}{AMD Instinct\texttrademark~MI300A}
\newcommand{\threeX}{AMD Instinct MI300X}
\newcommand{\threefive}{AMD Instinct\texttrademark~MI35X}
\newcommand{\threefifty}{AMD Instinct\texttrademark~MI350X}
\newcommand{\edp}{EDP$_{p}$}
\newcommand{\DESIGN}{PACER}
\newcommand{\SolutionName}{PACER}
\newcommand{\ExMeter}{DEM}
\newcommand{\CAC}{C$_{AC}$}
\newcommand{\ITwoR}{$I^2$R}
\newcommand{\perfLlamainf}{$-9\%\,$}
\newcommand{\perfLlamatrain}{$-15\%$} 
\newcommand{\perfSdxl}{}
\newcommand{\maxperf}{\Comment{XX}\%}
\newcommand{\geoperf}{\Comment{YY}\%}

\title{\SolutionName: GPU Excursion Management for ML Workloads}
\author{Rutwik Jain}
\authornote{Work done as intern at Advanced Micro Devices, Inc.}
\affiliation{%
 \institution{University of Wisconsin-Madison}
 \city{Madison}
 \country{USA}}
\email{rnjain@wisc.edu}

\author{Nathaniel Morris}
\affiliation{%
    \institution{Advanced Micro Devices, Inc.}
    \city{Austin}
    \country{USA}}
\email{Nathaniel.Morris@amd.com}

\author{Yiwei Jiang}
\affiliation{%
 \institution{University of Wisconsin-Madison}
 \city{Madison}
 \state{Wisconsin}
 \country{USA}
}
\email{jiang357@wisc.edu}

\author{Shidhartha Das}
\affiliation{%
  \institution{Advanced Micro Devices, Inc.}
  \city{Santa Clara}
  \state{California}
  \country{USA}
}
\email{Sid.Das@amd.com}

\author{Gary Kao}
\affiliation{%
  \institution{Advanced Micro Devices, Inc.}
  \city{Santa Clara}
  \state{California}
  \country{USA}
}
\email{Tzyy-juin.Kao@amd.com}

\author{Indrani Paul}
\affiliation{%
  \institution{Advanced Micro Devices, Inc.}
  \city{Santa Clara}
  \state{California}
  \country{USA}
}
\email{Indrani.Paul@amd.com}

\author{Shivaram Venkataraman}
\affiliation{%
\institution{University of Wisconsin-Madison}
\city{Madison}
\state{Wisconsin}
\country{USA}}
\email{shivaram@cs.wisc.edu}

\author{Srilatha Manne}
\affiliation{%
  \institution{Advanced Micro Devices, Inc.}
  \city{Santa Clara}
  \state{California}
  \country{USA}
}
\email{Srilatha.Manne@amd.com}

\author{Matthew D. Sinclair}
\affiliation{%
 \institution{University of Wisconsin-Madison}
 \city{Madison}
 \state{Wisconsin}
 \country{USA}
}
\email{sinclair@cs.wisc.edu}


\begin{abstract}
Variability is an inherent characteristic of modern GPU systems, arising from a combination of process variation, dynamic environmental conditions, and power management decisions. 
Prior work has characterized \emph{inter-GPU} variability and found consistent performance differences across GPUs.
However, we observe that modern workloads, particularly large language models (LLMs), exhibit significant \emph{intra-GPU}  variability. Their short-running kernels interact poorly with the coarse granularity of dynamic power management mechanisms, leading to unpredictable performance. 

A key source of this intra-GPU transient variability is power excursions, also referred to as \emph{electrical design power peaks} (\edp{}), where power temporarily exceeds the GPU's rated power limits. 
While power management mechanisms are designed to ensure reliable operation despite the presence of \edp{}, 
sustained power peaks cause throttling that detrimentally impacts overall application performance.
To the best of our knowledge, our work is the first to characterize the magnitude and frequency of \edp{}, showing that frequent, significant power overshoots are especially common in ML workloads, which combine high power draw, short kernels, and rapidly changing compute demands. 
To address this, we design and leverage hardware and firmware mechanisms to provide consistent power management behavior that reduces intra-GPU variability due to \edp{} and improves overall performance.
We evaluate our techniques across a range of popular workloads.
Our approach improves performance by 3--12\% over state-of-the-art GPU power management and comes within 3\% of an oracular solution.
\end{abstract}

%
%



\maketitle

\section{Introduction}
\label{sec:intro}

\begin{figure}[tb!]
  \centering
  \begin{subfigure}[b]{0.515\columnwidth}
    \centering
    \begin{tikzpicture}[font=\sffamily\scriptsize]
      \begin{axis}[
        width=\linewidth, height=0.73\linewidth,
        xlabel={Launch Year}, ylabel={TDP (Watts)},
        xmin=2016.2, xmax=2026.6,
        ymin=100, ymax=1560,
        xtick={2018,2020,2022,2024,2026},
        ytick={300,700,1100},
        clip=false,
        tick label style={font=\sffamily\fontsize{5}{6}\selectfont},
        label style={font=\sffamily\fontsize{6}{7}\selectfont},
        legend style={font=\sffamily\fontsize{5}{5.5}\selectfont,
                      inner sep=1pt, row sep=-2pt},
        legend image post style={mark size=1.2pt},
        every node near coord/.append style={
          font=\sffamily\fontsize{4.5}{5}\selectfont, inner sep=1pt},
        ]
        \addplot[annotated series, color=amdred, mark=*]
          table[col sep=comma, x=year, y=tdp, meta=product,
                discard if not={vendor}{AMD}] {data/tdp_trend.csv};
        \addlegendentry{AMD Instinct\texttrademark}

        \addplot[annotated series, color=nvgreen, mark=square*]
          table[col sep=comma, x=year, y=tdp, meta=product,
                discard if not={vendor}{NVIDIA}] {data/tdp_trend.csv};
        \addlegendentry{NVIDIA Data Center}
      \end{axis}
    \end{tikzpicture}
    \caption{}
    \label{fig:gpu-tdp}
  \end{subfigure}
  \hfill
  \begin{subfigure}[b]{0.445\columnwidth}
    \centering
    \includegraphics[width=\linewidth]{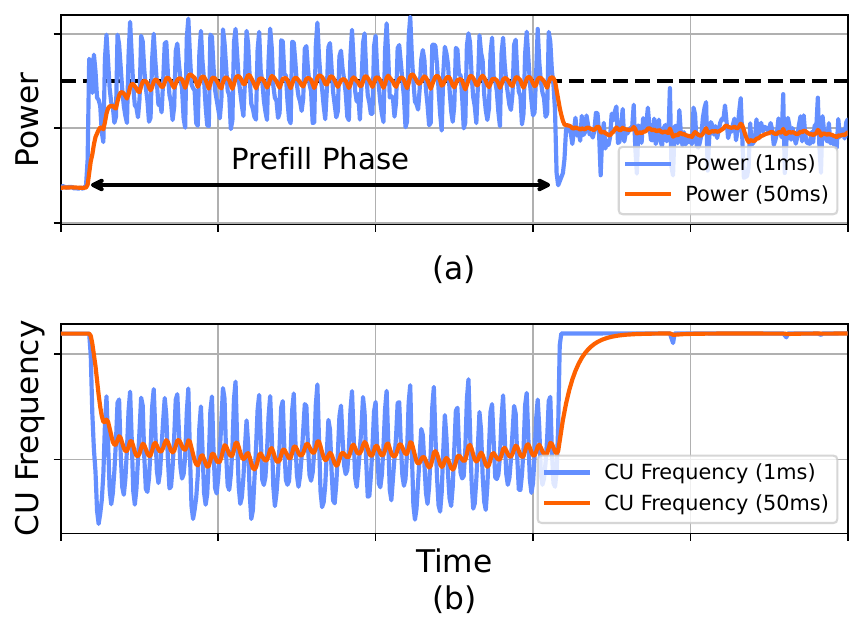}
    \caption{}
    \label{fig:llama-power-freq}
  \end{subfigure}
  \vspace{-1ex}
  \caption{(a) Thermal design power (TDP) of successive generations of AMD
  Instinct and NVIDIA datacenter GPUs by launch year.
  (b) Time series of power (top) and CU frequency set by the DPM (bottom)
  when running \qwen{} on an \threeX{} GPU. The dotted black line indicates the GPU's TDP.s}
  \Description{(a) Thermal design power of successive generations of AMD Instinct and NVIDIA data center GPUs by launch year. (b) Time series of power and CU frequency set by the DPM when running Qwen3-14B inference on an MI300X GPU; a dashed black line marks the GPU's TDP.}
  \label{fig:tdp-power-freq}
  \vspace{-2ex}
\end{figure}
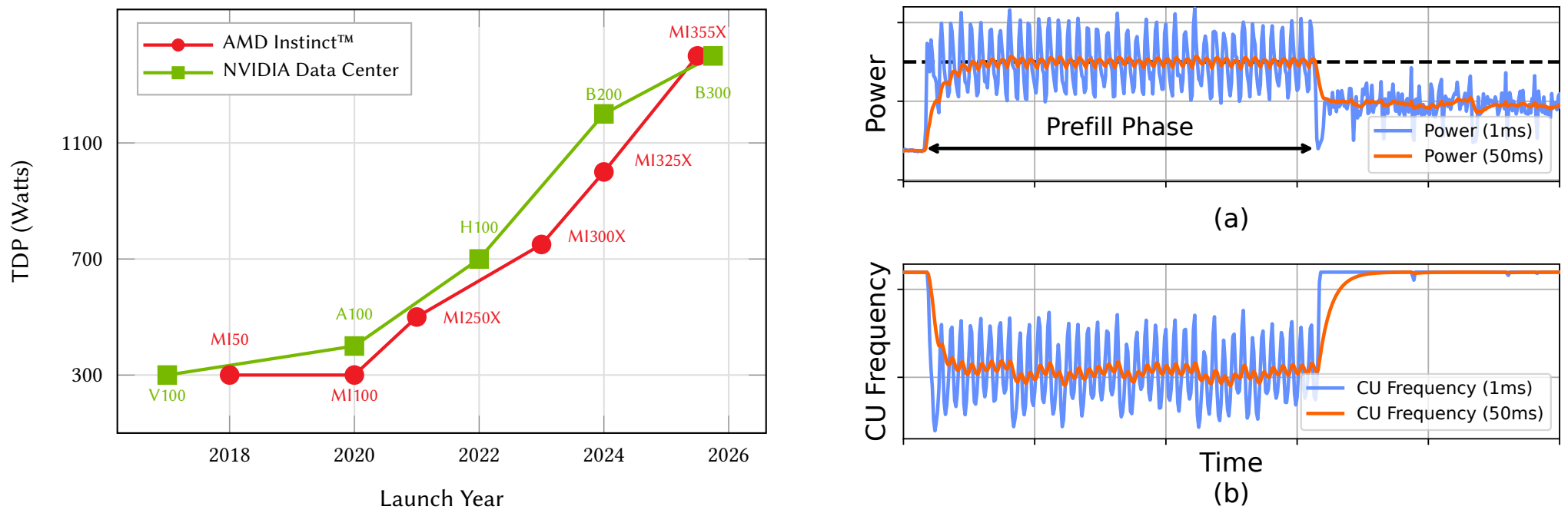

Voracious scaling demands in domains like graph analytics, high performance computing (HPC), and machine learning (ML) have led system developers to push technology limits with large, highly complex, heterogeneous systems that operate at previously unheard of power limits.
For example, recent systems such as \threeA{} and AMD Instinct\texttrademark~MI300X~\cite{cdna3, MI300A-ISCA2024} combine CPU/GPU solutions and stacked HBM using the latest in 2D and 3D chiplet technologies~\cite{jerger2014noc, kannan2015enabling, loh2015interconnect, SalehNaffziger2020-amdChipletPatent}.
Physically co-locating these components improves performance by shortening the distances between components and enables finer, higher bandwidth connections~\cite{MI300A-ISCA2024}.
Moreover, as shown in Figure~\ref{fig:gpu-tdp}, these GPUs dissipate up to 750 Watts with newer GPUs reaching 1400W~\cite{cdna4} and potentially 2300W~\cite{rubin-tdp}.
As a result, compute-intensive workloads such as ML are often power constrained on modern systems~\cite{amd-power, DeSensiDeMatteis2022-cloudPerfVar, JiangGarrigus2026-gpuLLMReliability, LengBuyuktosunoglu2021-gpuGuardband, Scogland2015-pwrPerspectives, sinha2022notall, TopcuKarabacak2025-gpuPowerPerfVar, YouXuan2024-gvarp, ZhongSultanov2025-uncorePowerWaste}.

To address these challenges, systems like the \threeX{} have robust power management mechanisms that dynamically adjust components like compute and data fabric frequencies (and, consequently, their power) to operate the system at the highest performance while staying within thermal design power (TDP) constraints.
The dynamic power manager (DPM) adjusts voltage and frequencies at short intervals, generally one to a few milliseconds~\cite{BharadwajDMBK23, power-stabilization, NathTullsen2015-crisp, Park-DVFS-Modeling, GreenGoverners, DPM-CCGRID}.
As long as application behavior is not highly variable across those time scales, most systems can successfully deliver the necessary performance within the device's power budget by managing voltage and frequency appropriately. 

However, together power constrained systems and dynamic power management mechanisms expose inherent \textbf{variability} among GPU systems.
Variability arises from various sources: process variation across chiplets, changing environmental conditions, and from the power-management decisions each GPU makes over time~\cite{DasOzdemir2007-processVar, InadomiPatki2015-scVar, sinha2022notall, TopcuKarabacak2025-gpuPowerPerfVar}.
We distinguish two forms of this variability, \textbf{inter-GPU} variability, which is the performance difference across GPUs running the same work, and \textbf{intra-GPU} or transient variability, which is variation over time when the same application or kernel runs on one GPU.
Previous work~\cite{sinha2022notall,Scogland2015-pwrPerspectives} characterized inter-GPU variability and found consistent variability \textit{across} GPUs, with very little \textit{intra-GPU} variability. 
However, in this work, we show that modern workloads, especially large language models (LLMs), have unique characteristics such as short kernels and variable compute requirements, which cause significant intra-GPU variability.

\begin{figure}[tb!]
   \centering
   \includegraphics[width=0.6\columnwidth]{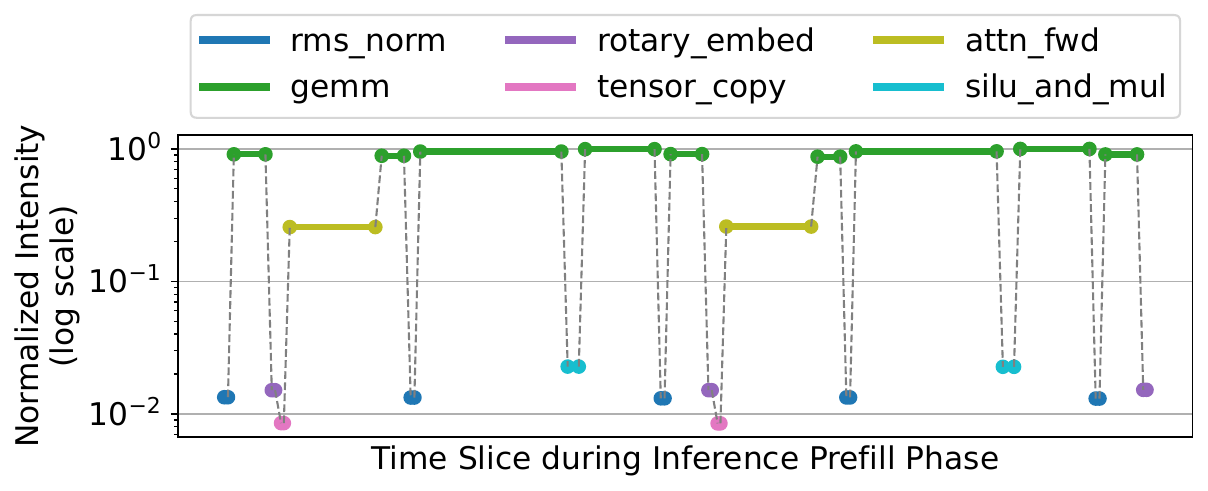}
   \vspace{-1ex}
   \caption{Time slice showing kernels from two transformer layers of the prefill phase for \llamainf{} with arithmetic intensity in operations/s, normalized between 0 and 1.}
   \Description{Time slice showing kernels from two transformer layers of prefill phase for \llamainf{} with arithmetic intensity in operations/s, normalized between 0 and 1.}
   \label{fig:arith-intensity}
   \vspace{-2ex}
\end{figure}

Applications such as LLMs not only push the TDP limit of these systems, they also have rapidly fluctuating behavior varying between compute dense, high power phases and lower power phases at microsecond timescales. 
Figure~\ref{fig:arith-intensity} shows how the arithmetic intensity of the kernels varies within two Transformer layers of the prefill phase of LLM Inference, revealing fluctuations spanning over two orders of magnitude across kernels with sub-millisecond durations.
The timescale is far shorter than the DPM's adjustment interval (kernel fusion's impact discussed further in Section~\ref{subsec:exbehavior-gemms}).
Moreover, modern ML workloads increasingly utilize diverse forms of parallelism and concurrency, from model and expert-level parallelism~\cite{MegatronLM,HwangWei2025-moe, Mixtral} to computation layer and kernel-level micro-batching~\cite{HuangCheng2019-gpipe,Orion-interleaving}. 
These techniques are essential for scaling models beyond single-device memory capacity, improving hardware utilization, and reducing per-token serving costs.
While these solutions improve performance, the microsecond timescales at which application characteristics change clash with the longer intervals for GPU dynamic power management solutions.
The slower reactivity of the DPM relative to application behavior impacts transient kernel variability, and it also results in power peaks when the GPU is already operating near its power limits and the application rapidly switches between low and  high activity phases~\cite{chasapis2016runtime, ChasapisMoreto2019-powerEfficJobSched, DasOzdemir2007-processVar, InadomiPatki2015-scVar, blackwell}.

To understand how frequently peak GPU power exceeds TDP for modern workloads, Figure~\ref{fig:llama-power-freq} shows how GPU power and compute unit (CU)\footnote{Without loss of generality, we use AMD GPU terminology.  NVIDIA GPUs have equivalent terminology (e.g., SMs vs. CUs).} frequency vary for \qwen{}~\cite{dubey2024llama3herdmodels}, run using vLLM~\cite{kwon2023efficient}.
The orange lines indicate power and DPM set frequency, respectively, averaged over 50 millisecond (ms) intervals.
The blue lines 
show the same metrics at $\approx$1 ms intervals.
As prior work observed~\cite{patel-2024-splitwise, LLM-ASPLOS}, inference power in Figure~\ref{fig:llama-power-freq} (top) varies between the prefill and decode phases: prefill generally dissipates near TDP power, and decode operates below the TDP.
Due to power constraints, the average CU frequency is lower during the prefill phase than during decode.
By adjusting the CU frequency, the DPM optimizes performance under power constraints.
Moreover, although power looks steady 
when averaged across the larger 50 ms intervals (orange lines)~\cite{YangAdamek2024-gpuPowerMeas}, smaller duration averages show the \edp{} swinging beyond 
TDP (blue lines).
These peaks occur frequently during an LLM's prefill phase and less frequently during the decode phase.
Each peak is followed by a frequency correction, so the same kernel does not see a stable frequency from one instance to the next.
The lack of a stable operating frequency is the main cause of intra-GPU variability.
More generally, we find that \edp{} occurs in applications with short kernels and significant compute variability across adjacent kernels (discussed further in Section~\ref{sec:exbehavior}).

While previous works have observed \edp{} (discussed in Section~\ref{sec:previous}), a broader study is needed to understand how prevalent they are across applications, and how \edp{} impacts application performance.
Thus, we first perform a \textbf{comprehensive analysis of \edp{} on GPUs across modern graph analytics, HPC, and ML workloads to identify which application characteristics result in power overshoots}.
Our results show that \edp{} events are triggered at especially high rates in ML workloads like \qwenmodel{} and \llamainf{}.
Moreover, \edp{} will likely become even more challenging for future GPUs with further chiplet integration and higher TDPs.
We also show that \textbf{\edp{} can be both beneficial and detrimental -- properly managed \edp{} enable applications to computationally sprint~\cite{ComputationalSprinting,Morris17-EarlySprint,MorrisRenganathan2016-compSprintSW,FanZahedi2016-compSprintGame} while excessive high-magnitude \edp{} incur performance loss due to throttling mechanisms}.

We use the insights from this characterization to manage excursions so that \edp{} stays in the range that preserves computational sprinting, without the throttling that produces intra-GPU variability. 
In hardware, \edp{} clamping reduces peak magnitude, at a cost in performance. 
In firmware, a graphics frequency cap together with a fixed memory-clock state narrows the range over which the DPM can move, so successive kernels see more repeatable voltage and frequency decisions.
We evaluate both mechanisms on an experimental system built atop production \threeX{} GPUs, with no changes to the applications.
Frequency limits improve prefill performance and reduce transient kernel variability for single-GPU inference.
The same caps, chosen from offline profiling and applied based on dynamic request characteristics, improve multi-GPU online serving.
We \textbf{analyze these \edp{}--performance tradeoffs across popular ML workloads for both firmware and hardware management}. 
Overall, our solutions improve performance over state-of-the-art, production-grade power management by 3--12\% on \threeX{} GPUs. 
Notably, our designs provide near-best-case performance: within 3\% of an oracle power management mechanism.

\section{Background}
\label{sec:background}

\edp{} events occur when current consumption exceeds a certain threshold. 
\edp{} can result in 
current increases that 
can drain package decoupling capacitors and push voltage regulators out of compliance, potentially triggering system protection mechanisms like power throttling.
Power overshoots have thermal and reliability implications~\cite{BakerVikas2024-powerCycling, HoffmanSchmitt2023-powerSwingsLowTemp, intel-vrm-evrd10, MarieCougo2024-thermalReliability, OnoderaShishido2024-solderPowerCycleLifetime, Savino2025-powerForAI, SchwabeSeltner2024-reliability, Wang-Reliability-PowerElectronics, Zhang-EMI-PowerElectronics, ZhangHeimler2026-powerCycling} that, if not managed, may cause catastrophic hardware failures such as physical damage to VRs or even catching fire~\cite{cadence-powerIssues, GuoLu2023-circuitDamage, PowerReliability, KovacevicKolar2015-modellingSemiconductors, Wallossek2025-autoDownReg}.
\begin{table}[tb!]
  \caption{(a) \edp{} limits from the OCP specification and (b) the metrics
  and terminology used in this paper.}
  \vspace{-1ex}
  \centering
  \begin{subtable}[t]{0.42\columnwidth}
    \centering
    \caption{OCP~\cite{ocp-spec} specification for \edp{}.}
    \label{tab:ocpspec}
    \scriptsize
    \begin{tabular}{@{}lll@{}}
    \toprule
    { \textbf{\edp}} & { \textbf{Duration}} & { \textbf{Management Solution}} \\ \midrule
    2$\times$ TDP    & Up to 20$\mu$s  & HW       \\
    1.7$\times$ TDP  & Up to 2ms       & HW or SW \\
    1.5$\times$ TDP  & Up to 5ms       & SW       \\
    1.25$\times$ TDP & Up to 10ms      & SW       \\
    1.1$\times$ TDP  & Up to 20ms      & SW       \\
    1$\times$ TDP    & Indefinitely    & SW       \\ \bottomrule
    \end{tabular}
  \end{subtable}
  \hfill
  \begin{subtable}[t]{0.56\columnwidth}
    \centering
    \caption{Metrics and terminology glossary.}
    \label{tab:metrics}
    \scriptsize
    \begin{tabular}{@{}lll@{}}
  \toprule
  \multicolumn{2}{l}{\textbf{Metric}} & \textbf{Description} \\ \midrule
  EDP$_{p}$     & \begin{tabular}[c]{@{}l@{}}Electrical Design\\ Power peaks\end{tabular} & \begin{tabular}[c]{@{}l@{}}Short time-scale power overshoots\\ or spikes above TDP.\end{tabular} \\
  $C_{AC}$ & \begin{tabular}[c]{@{}l@{}}Active switching\\ capacitance\end{tabular}    & \begin{tabular}[c]{@{}l@{}}GPU capacitive activity that drives\\ dynamic power consumption.\end{tabular}     \\ \bottomrule
\end{tabular}

  \end{subtable}
  \label{tab:ocpspec-metrics}
  \vspace{-2ex}
\end{table}

These challenges show how \edp{} events require a robust and reliable hardware protection mechanism to ensure over-current protection and safe operation.
\Highlight{Given the catastrophic nature of the problem, \edp{} management must have a dependable backstop to ensure reliable operation.}
The Open Accelerator Infrastructure (OAI) OCP Accelerator Module specification provides an exemplar guide for the voltage regulator (VR) electrical design to handle power overshoots~\cite{ocp-spec}.
The OCP specification defines the tolerable peak power of the system as a function of the socket's TDP.
For example, Table~\ref{tab:ocpspec} outlines the board rail of a hypothetical 700W system.
For instance, this system can draw up to $2\times$ the TDP power limit as long as that peak does not exceed 20 $\mu$s.
Meeting this specification, and not exceeding TDP over a period of 1 second~\cite{ocp-spec, power-stabilization}, ensures the VR will be reliable and operate as specified.
Other GPU products are designed for a more relaxed specification with longer peak power tolerances (e.g., 50 ms duration).
However, the more relaxed the specification, the higher the cost: additional board capacitance requirements and likely a higher power VR, which impacts the next level of the power delivery hierarchy, accruing further costs up the power delivery chain to the rack and potentially beyond~\cite{power-stabilization}.

\edp{} response depends on how long the peak is allowed to last and the DPM's capabilities.
In modern systems, some of the listed overshoot time constants in Table~\ref{tab:ocpspec} (e.g., $2\times$TDP) require hardware solutions to manage at the microsecond time scale. 
However, firmware management solutions that operate at millisecond scales can utilize dynamic voltage and frequency scaling (DVFS) to account for Table~\ref{tab:ocpspec}'s longer durations (or those in prior work~\cite{power-stabilization}).

Every fail-safe \edp{} management mechanism must meet the VR's specified requirements.
The management solution \textbf{cannot} fail, so the resulting solution has built in margins to deal with power measurement inaccuracies, delays between measurement and evaluation, and response capabilities.
For instance, a hardware solution that samples power from the board power supply will have very accurate data.
However, there may be a relatively long latency to receive the data, whereas a digital power estimation solution can be computed quickly but will have some estimation error.
The \edp{} response may involve architectural solutions such as temporarily shutting down some sub-modules, reducing frequency~\cite{AdaptiveClocking, AdaptiveClocking-Grenat}, or reducing both voltage and frequency (DVFS).
Whatever solution is selected, it must be responsive enough to deal with the large overshoots that can rapidly occur in a highly parallel, highly synchronized, high switching capacitance system like a GPU.
For one or more of these reasons, an \edp{} management mechanism will likely have significant margining to ensure VR protection (e.g. conservative \edp{} thresholds to trigger throttling) .
If the frequency of \edp{} is limited, the management solution's efficiency will not matter.
However, frequent, high-magnitude \edp{} events that trigger the management solution and power throttling will degrade application performance.

\section{\texorpdfstring{\edp{}}{EDP} Characterization}
\label{sec:exbehavior}

We characterize \edp{} behavior across diverse applications and compare the magnitude and frequency of \edp{}.
We use the \threeX{} GPU~\cite{cdna3} which have a TDP of 750W. 
As defined in Section~\ref{sec:intro}, \edp{} refers to power peaks beyond the GPU's TDP limit.
Table~\ref{tab:metrics} defines the metrics and terminology, including \edp{}, that we use in the remainder of this paper.
We measure GPU power using AMD GPU hardware counters, which provide 1ms granularity of power measurements~\cite{rsmi-api}.
Specifically, we profile the GPU's Total Board Power\footnote{Total Board Power includes the GPU and HBM power as well as any power delivery overheads associated with the board power supply.} and report measurements relative to a TDP of 750W.
To ensure consistent results, we repeat all experiments three times and present either the median results or the distribution across all trials.
We show cumulative distributions of this measured power, specifically showing the power distribution at and over TDP.
Due to the 1 ms averaging of power, we cannot track smaller duration events with high amplitude, such as $2\times$TDP in Table~\ref{tab:ocpspec}.
Any \edp{} events shorter than 1 ms in duration are dampened and absorbed into the 1 ms average power results.
Nevertheless, the fact that we still observe \edp{} even at this coarser granularity highlights the importance of the problem.

\begin{scriptsize}
\begin{table}[tb!]
\centering
\caption{Workloads used in characterization.}
\label{tab:bmks}
\setlength{\tabcolsep}{4pt}
\begin{tabular}{@{}lll@{}}
    \toprule
    \textbf{Workload} & \textbf{Domain} & \textbf{Dataset/Config} \\ \midrule
    Qwen3-14B~\cite{qwen3} inference & ML (LLM) & bsz 1, ip 4k, op 128 \\
    LLaMA3-8B~\cite{touvron-2023-llama,dubey2024llama3herdmodels} inference & ML (LLM) & bsz 1/4/8/16, ip 1k/4k, op 128 \\
    LLaMA3-70B~\cite{touvron-2023-llama,dubey2024llama3herdmodels} inference & ML (LLM) & bsz 1/4, ip 1k/4k, op 128 \\
    Mixtral-8x7B~\cite{Mixtral} inference & ML (LLM) & bsz 1/4, ip 1k/4k, op 128 \\
    Stable Diffusion~\cite{rombach2021highresolution,podell2023sdxlimprovinglatentdiffusion} & ML (Diffusion) & bsz 1 \\
    \midrule
    MILC~\cite{MILC} & HPC & su3\_rhmd\_hisq, $24\times24\times24$ \\
    LULESH~\cite{LULESH:prog,LULESH:spec,LULESH2:changes,IPDPS13:LULESH} & HPC & n 400, i 10 \\
    \midrule
    PageRank~\cite{WangPan2017-gunrock} & Graph Analytics & \Edit{indochina, uk-2005} \\
    \Edit{Neighbor Sampling}~\cite{PACT-Skywalker,WebGraph-uk-1,WebGraph-uk-2} & \Edit{Graph Analytics} & \Edit{indochina, uk-2005, Orkut} \\
    \Edit{Random Walk}~\cite{PACT-Skywalker,WebGraph-uk-1,WebGraph-uk-2} & \Edit{Graph Analytics} & \Edit{indochina, uk-2005, Orkut} \\
    \Edit{K-Core}~\cite{PACT-Skywalker,WebGraph-uk-1,WebGraph-uk-2} & \Edit{Graph Analytics} & \Edit{indochina, uk-2005, Orkut} \\
\bottomrule
\end{tabular}\vspace{-2ex}
\end{table}
\end{scriptsize}

Power variability occurs due to a combination of power management directives and application characteristics.
The goal of power management firmware (PMFW) is to achieve the best performance possible for a given application.
PMFW increases GPU clock frequency and the associated voltage until either maximum board power or maximum frequency is reached.
This results in higher frequencies when running medium- or light-density compute operations and lower frequencies when running compute intensive applications with high switching capacitance and high power.
Similar to many CPU power management algorithms, GPU PMFW uses past behavior as an indicator of future requirements.
If application behavior does not change or changes slowly over many milliseconds, the GPU PMFW can easily adjust DVFS states to adapt to application characteristics and hold power steady at the board power limit~\cite{BharadwajDMBK23, amd-power, CoplinBurtscher2016-gpgpuPower, FanCosenza2019-predGPUFreq, GeVogt2013-dvfsKepler, MeiYung2013-gpuDVFSMeasure, NathTullsen2015-crisp, sinha2022notall}.
However, as Figure~\ref{fig:arith-intensity} shows, the fluctuating load in ML applications in combination with a firmware DPM that operates on a larger time scale than microseconds can produce power overshoots due to a mismatch between the operating frequency and the switching capacitance.
\Highlight{Thus, while previous work has observed some \edp{} for specific workloads~\cite{power-stabilization, Patel-CAL}, a broader study is necessary to understand the extent and frequency of \edp{} events in modern systems.}

\subsection{Modern GPU Workload Characteristics}
\label{subsec:exbehavior-hpcml}

To characterize \edp{} across diverse workloads and compare their behavior with LLMs, we profiled the power consumption and power peaks for popular graph analytics, HPC, and ML workloads from Gunrock~\cite{WangPan2017-gunrock}, \Edit{Skywalker~\cite{PACT-Skywalker}}, CORAL-2 and OLCF-6~\cite{MILC, LULESH:prog, IPDPS13:LULESH, LULESH2:changes, coral2}, and MLPerf~\cite{MattsonCheng2019-mlperfTrain, Reddi2020mlperf-Infer}.
These workloads provide wide coverage across a range of access patterns and compute intensities while also ensuring they are representative of workloads driving the requirements for future systems.
For all applications, we either selected inputs that utilized the GPUs well or examined multiple inputs (e.g., \llamainf{} with different batch sizes).
Table~\ref{tab:bmks} summarizes these workloads and their inputs.
\begin{figure}[tb!]
    \centering
    \includegraphics[width=0.8\linewidth]{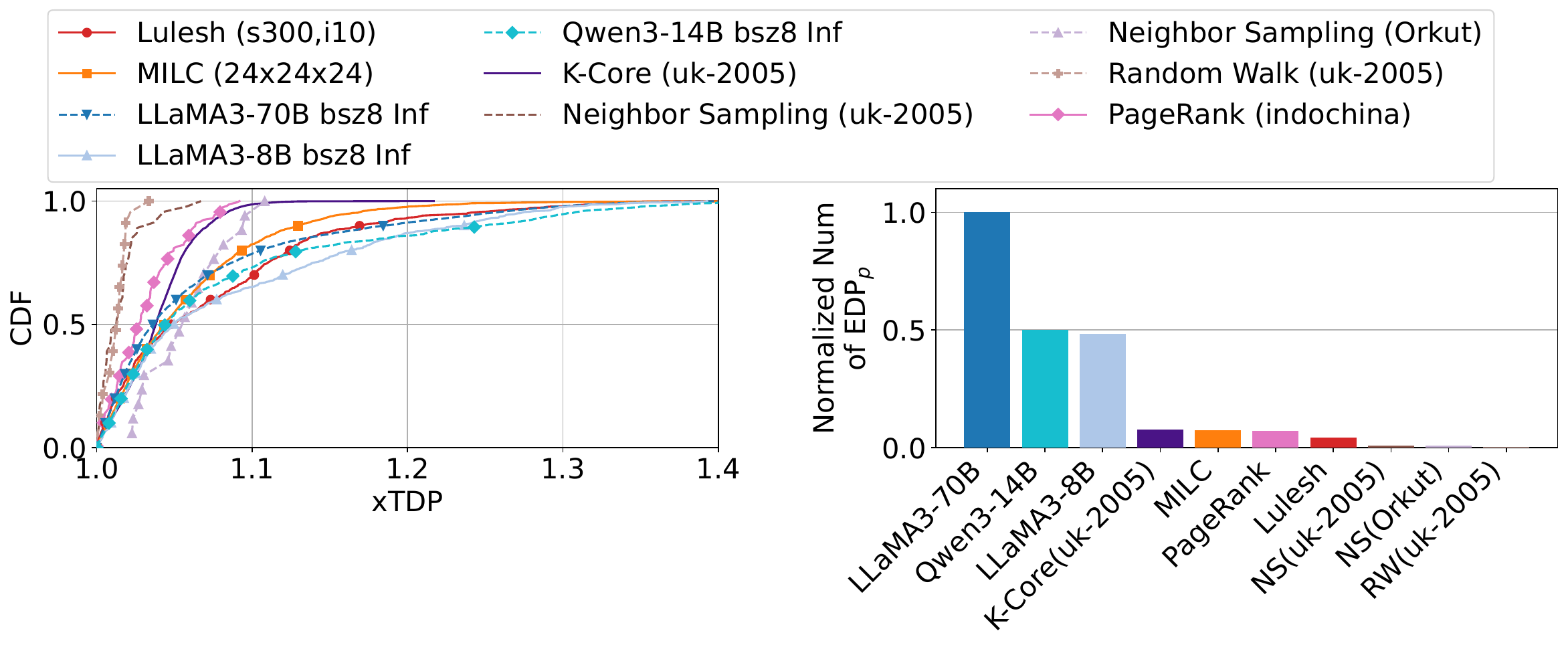}
    \vspace{-2ex}
    \caption{Cumulative Distribution (CDF) of \edp{} above TDP for HPC, ML, and graph analytics applications on a single \threeX~GPU.}
    \Description{Cumulative Distribution (CDF) of \edp{} above TDP for HPC, ML, and graph analytics applications on single \threeX~GPU.}
    \label{fig:excursions-mi300x}
    \vspace{-3ex}
\end{figure}

Figure~\ref{fig:excursions-mi300x} shows the cumulative distribution function (CDF) of \edp{} values for these workloads on an \threeX{}~GPU, as well as the number of \edp{} events normalized to the application with the most \edp{} (LLaMA3-70B inference).
These overshoots are not unique to the system being analyzed: \edp{} of up to 500W over TDP have been observed on other accelerators~\cite{LLM-ASPLOS, power-stabilization}.
Interestingly, all ML applications (LLaMA2 fine-tuning, LLaMA3 inference, and Stable Diffusion) exhibit very similar \edp{} CDFs regardless of batch size or application: $\approx$80\% of their peaks are between 1.0$\times$ and 1.2$\times$ of TDP, but their remaining peaks exhibit a long tail, and some of them can be quite significant: max 1.7$\times$, although this max happens $<$ 1\% of the time.
\Edit{Of the graph analytics and HPC workloads (PageRank, Neighbor Sampling, Random Walk, K-Core, Lulesh, and MILC), only Lulesh shows a significant portion of \edp{} above 1.2$\times$TDP.
All graph analytics workloads and MILC have much less variable behavior: over 90\% of their \edp{} is below 1.2$\times$TDP.
We also ran Neighbor Sampling, K-Core, and Random Walk workloads on the indochina dataset, but did not observe any excursions for them. 
Therefore, Figure~\ref{fig:excursions-mi300x} does not include these workload-input combinations.}
Moreover, although the relative fraction of \edp{} shows that workloads like Lulesh might still see high \edp{} magnitudes, these overshoots occur infrequently.
Conversely, ML workloads 
have frequent and high magnitude \edp{} events.
Therefore, we focus our characterization and eventual evaluation of \edp{} management solutions on ML workloads.

\subsection{Inter-Kernel ML Workload Characteristics}
\label{subsec:exbehavior-ml}

ML workloads are characterized by extremely short duration kernels, on the order of a few 100s of microseconds to a few milliseconds, and high-power variability between adjacent kernels~\cite{JainTran2024-pal, sinha2022notall, YouXuan2024-gvarp}.
Short kernels make it difficult to effectively modify DVFS state from one kernel to the next~\cite{DuttaAdhinarayanan2018-gpuPowerPredML, FanCosenza2019-predGPUFreq, MishraKhare2016-gpuDVFSEval, NathTullsen2015-crisp, NugterenVanDenBraak2014-rooflineGPUDVFS, TangWang2019-gpuDVFSDL}.
In addition, adjacent kernels that transition from lower to higher compute density mean that a DVFS state selected for a lower power kernel may result in overshoots for successive kernels with high compute intensity.

\begin{figure}[tb!]
    \centering
    \begin{subfigure}{0.47\linewidth}
        \includegraphics[width=\linewidth]{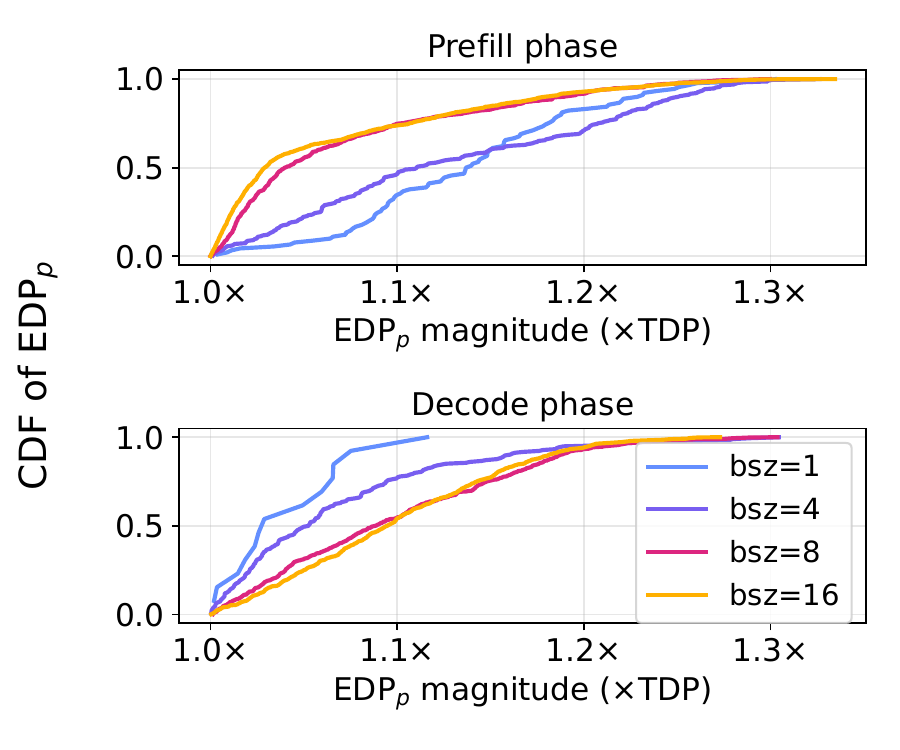}
        \caption{}
        \label{fig:excursions}
    \end{subfigure}
    \hfill
    \begin{subfigure}{0.47\linewidth}
        \includegraphics[width=\linewidth]{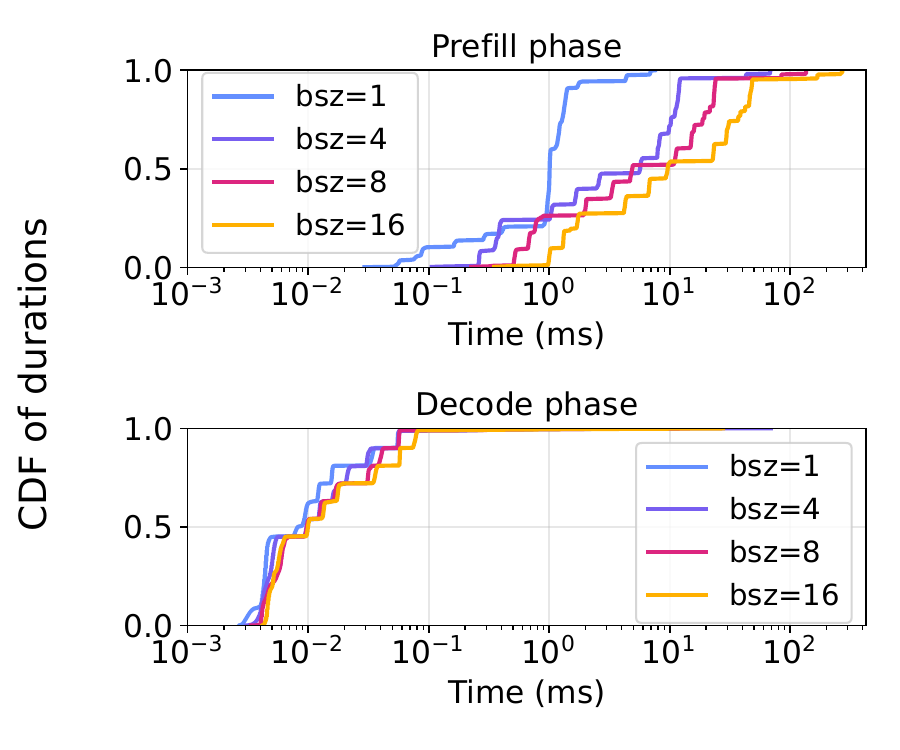}     
        \caption{}
        \label{fig:kernel-dur-cdf}
    \end{subfigure}
    \vspace{-2ex}
    \caption{(a) \edp{} distribution and (b) cumulative distribution of kernel durations for prefill and decode phases of \llamainf{} with varying batch sizes (bsz) on an \threeX~GPU.}
    \Description{(a) \edp{} distribution and (b) cumulative distribution of kernel durations for prefill and decode phases of \llamainf{} with varying batch sizes (bsz) on an \threeX~GPU.}
    \vspace{-2ex}
\end{figure}

We use \llamainf{} to highlight the relationship between short runtime kernels and \edp{} characteristics.
Figures~\ref{fig:excursions} and \ref{fig:kernel-dur-cdf} show the CDF of overshoot magnitudes and kernel durations, respectively, for \llamainf{} with varying batch sizes ranging from 1 to 32. 
Both figures average their results over three trials. 
\edp{} is separated into prefill and decode phases. 
Prefill not only has the majority of overshoots, but it also has larger magnitude peaks. 
In prefill, small batch sizes produce larger \edp{} due to short kernel durations (Figure~\ref{fig:kernel-dur-cdf}) resulting in high frequency fluctuations between high- and low power kernels. 
As the prefill batch size increases, the kernels increase in duration, enabling the DPM to adjust the DVFS state appropriately.
In the case of decode, the opposite occurs -- higher batch sizes result in larger peaks. 
Larger batch sizes in decode also increase kernel duration but not to the same extent as prefill. 
However, larger batch sizes increase decode power so that it begins to operate near the GPU's TDP, increasing the likelihood of \edp{} due to application load fluctuations.

\begin{figure}[tb!]
    \centering
    \begin{subfigure}{0.4\linewidth}
        \centering
        \includegraphics[width=0.8\linewidth]{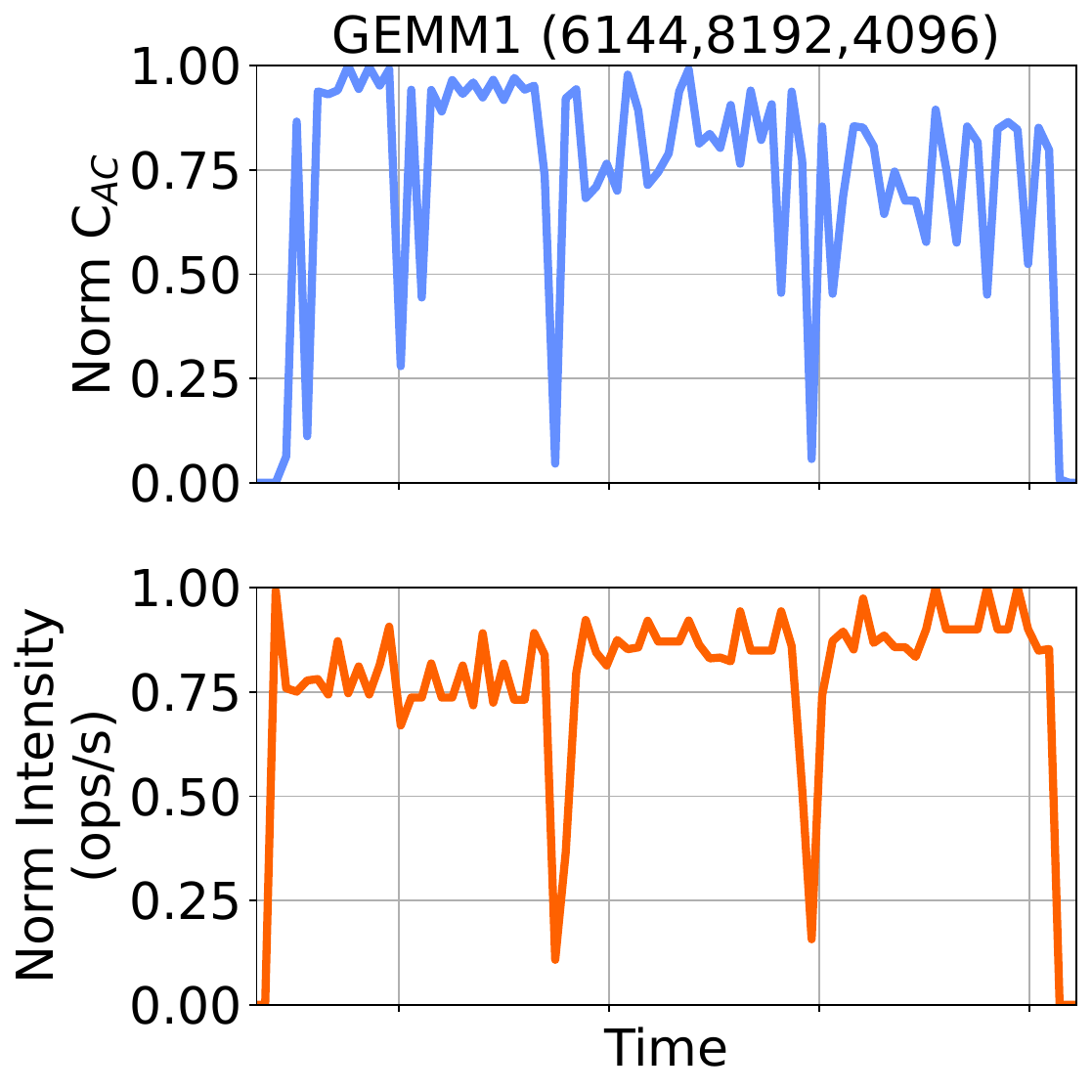}
        \caption{}
        \label{fig:gemm1}
    \end{subfigure}
    \begin{subfigure}{0.4\linewidth}
        \centering
        \includegraphics[width=0.8\linewidth]{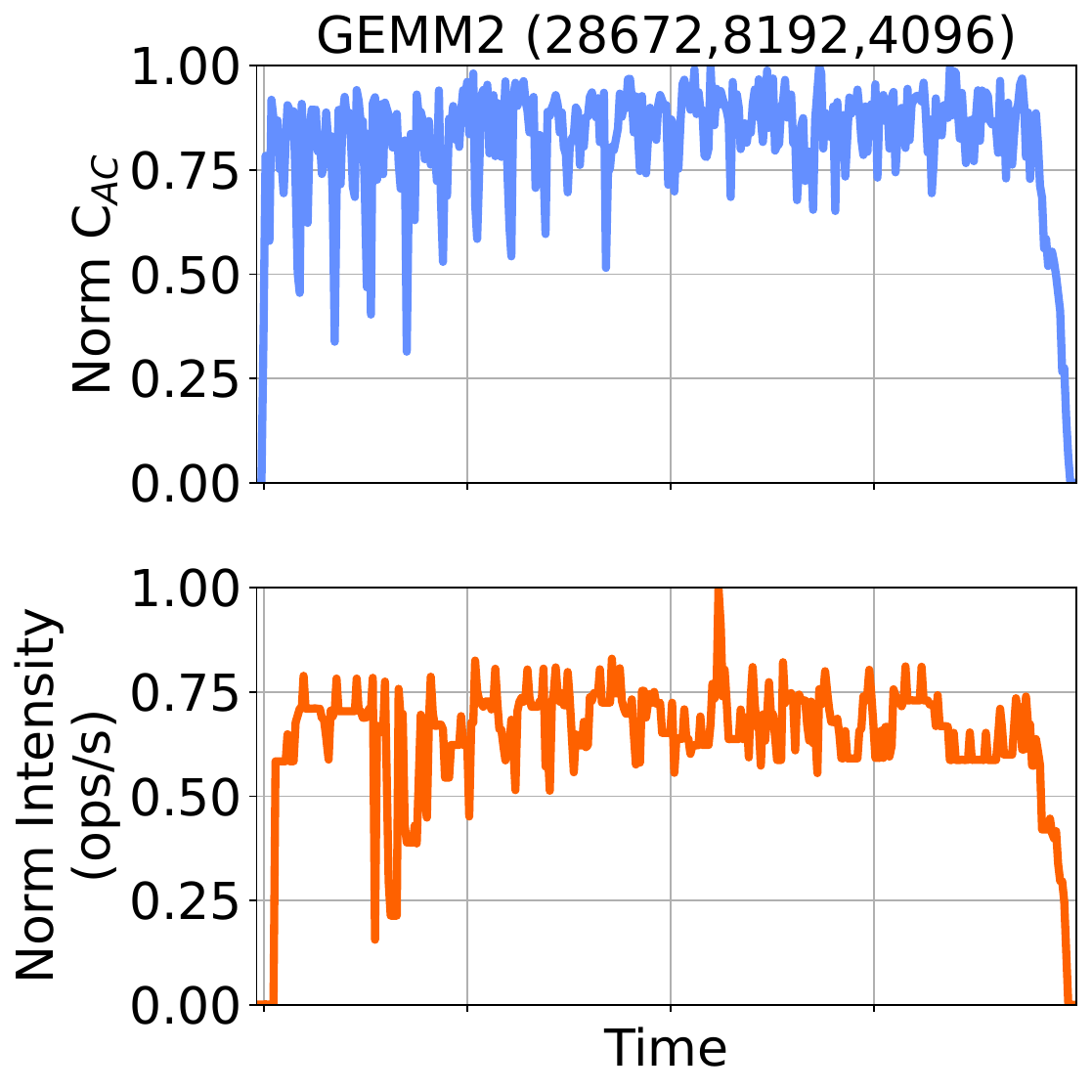}
        \caption{}
        \label{fig:gemm2}
    \end{subfigure}
    \vspace{-2ex}
    \caption{Intra-kernel variation in normalized \CAC{} and arithmetic intensity in operations/s for GEMMs within \llamainf{}.}
    \Description{Intra-kernel variation in normalized \CAC{} and arithmetic intensity in operations/s for GEMMs within \llamainf{}.}
    \label{fig:gemm-var}
    \vspace{-2ex}
\end{figure}

\subsection{Intra-kernel ML Workload Characteristics}
\label{subsec:exbehavior-gemms}

Inter-kernel GPU power variability is common since natural shifts in activity and power may occur at kernel boundaries, leading to \edp{}.
As highlighted in Table~\ref{tab:metrics}, we quantify activity using capacitive switching or \CAC{}.

While shorter kernels exacerbate \edp{} due to fine-grained fluctuations in activity, making kernels longer through techniques like kernel fusion~\cite{Appleyard2016-optRNNGPU, FilipovivcMadzin2015-blasCUDAFusion, LiZheng2020-horizFus, cudnn, WangLin2010-gpuKernelFusion} or megaKernels~\cite{stanford-megaKernel2, stanford-megaKernel1} does not resolve the problem.
Longer kernels also incur intra-kernel power variability.
For example, the computation in fused kernels (e.g., a compute-intensive GEMM and a memory-intensive elementwise operation) differs significantly.
Moreover, even within the GEMM phase, compute requirements can vary.
To illustrate this, consider GEMM kernels within LLM inference.
At a high level, GEMMs appear uniform in their behavior, as shown in Figure~\ref{fig:arith-intensity}.
However, tail effects are an inherent aspect of GEMM tiling on modern GPU architectures due to the interaction between matrix dimensions $(M, N, K)$ and the hardware’s organization of compute resources.

When a GEMM kernel is tiled, the input matrices are partitioned into blocks (or tiles), each assigned to a workgroup (WG).
These WGs are further decomposed into wavefronts (WFs), which are scheduled across the available CUs.
The number of threads per WF and the mapping of WGs to CUs are determined by the architecture, but the sizes of $M$, $N$, and $K$ are dictated by the application and rarely align perfectly with the chosen tile sizes~\cite{Rammer, Streamk, PatiAga2025-goldyloc}.
As a result, the final tiles along each dimension -- the tails -- are almost always partial.
Figures~\ref{fig:gemm-var}a and~\ref{fig:gemm-var}b show arithmetic intensity in TFLOP/s for two GEMMs within \llamainf{} when running on \threeX{} -- GEMM1 with $(M, N, K)$ dimensions $(6144, 8192, 4096)$ and GEMM2 with $(M, N, K)$ dimensions $(28672, 8192, 4096)$ respectively.
Both GEMMs show periodic dips in arithmetic intensity: points within the kernel where one set of WGs completes while the next set of WGs is launched. 
For the smaller GEMM1, work is divided into three sets of WGs that can each run concurrently on the GPU's 304 CUs.\footnote{For this hipBLAS GEMM, number of WGs is: $\frac{M}{m_{tile}}\times\frac{N}{n_{tile}} = 877$ and $\frac{877}{304 \text{CUs}} = 2.88$. The tail still utilizes 88\% of CUs. Similar calculations for the other GEMM yield 23.5, which means the 24th set of WGs will utilize only about half the GPU.}
Here, the tail set of WGs has roughly the same arithmetic intensity as the previous two.
Conversely, GEMM2 shows a long tail where arithmetic intensity steps down.
In GEMM2 the tail WGs process fewer elements than the regular tiles, causing underutilization. 

This mismatch is not simply a software artifact, but a direct consequence of the discrete mapping between the logical decomposition of the GEMM workload and the GPU's physical resources. 
This quantization effect is fundamental to the architecture. 
Even with careful tile size selection to match the preferred WF and WG configurations, the 
nature of matrix dimensions means that the last few tiles will often be incomplete.
This reduction in compute requirements, combined with natural process variations across dies within a GPU that produces varying completion times across the SoC, reduces power at the end of a GEMM computation. 
The DPM detects the lower power and increases frequency accordingly, resulting in variability even within a kernel.
Figure~\ref{fig:gemm-var}'s arithmetic intensity of GEMMs reflects this: the change at the end of the previous set of WGs allows the DPM to increase the frequency for a short duration before the next set of WGs is launched.
This also means that the same GEMM computations may experience different power fluctuations and \edp{}, depending on what kernel precedes it, and how it interacts with the DPM.



\section{\texorpdfstring{\edp{}}{EDP} Management}
\label{sec:design}

Section~\ref{sec:exbehavior}'s characterization shows that ML applications exacerbate physical GPU constraints with high compute requirements and highly variable application behavior.
Moreover, they can also impact performance via throttling if they occur too often and require significant management~\cite{mlperf25}.
However, we haven't quantified the impact of \edp{} on performance.
Accordingly, we look at the following research questions to evaluate \edp{} and \edp{} management:\\

RQ1. How does \edp{} impact performance?
RQ2. How can we manage \edp{} to improve performance?

We evaluate the hardware \edp{} for inference workloads (Section~\ref{sec:design-hw}) to show that lower-magnitude \edp{} is beneficial for performance, while higher-magnitude \edp{} needs to be mitigated. 
Accordingly, we design a dynamic runtime approach to mitigate \edp{} by managing frequency limits for the compute and memory clocks. 
Overall, across a range of popular ML workloads, versus a state-of-the-art, production grade GPU power management scheme, our solution improves performance by up to 12\% on production systems (which is within 3\% of an oracular solution).

\subsection{RQ1: How does \texorpdfstring{\edp{}}{EDP} impact performance?}
\label{sec:design-hw}

\begin{figure}[tb!]
    \centering
    \begin{subfigure}[b]{0.66\textwidth}
        \centering
        \includegraphics[width=\linewidth]{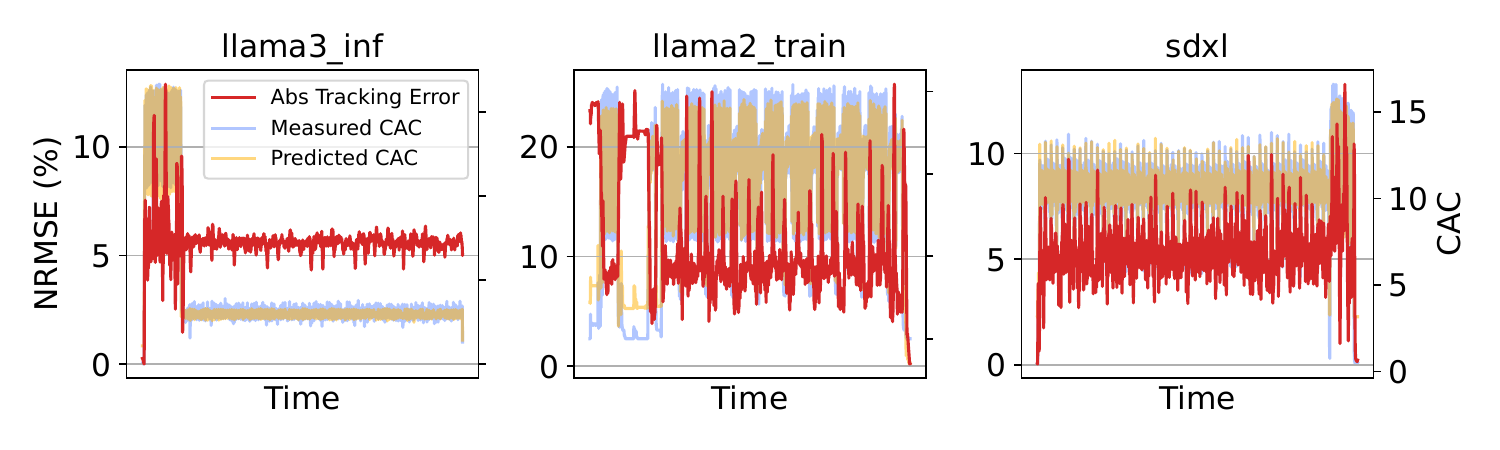}
        \caption{}
        \label{fig:cac-timeseries}
    \end{subfigure}
    \hfill
    \begin{subfigure}[b]{0.30\textwidth}
        \centering
        \begin{scriptsize}
        \begin{tabular}{@{}ll@{}}
            \toprule
            \textbf{Workload} & \textbf{NRMSE (\%)}\\
            \midrule
            \llamainf{} & 5.7 \\ 
            LLaMA2 training & 12.3 \\ 
            Stable Diffusion (SDXL) & 5.6 \\ 
            \bottomrule
        \end{tabular}
        \end{scriptsize}
        \caption{}
        \label{fig:cac-nrmse}
    \end{subfigure}
    \vspace{-2ex}
    \caption{(a) Timeseries showing absolute prediction error (left Y) and normalized measured and predicted \CAC{} (right Y) for \llamainf{}, LLaMA2 training, and Stable Diffusion (sdxl) and (b) \% Normalized Root Mean Squared Errors (NRMSE) for \CAC{} predictions for each workload.}
    \label{fig:cac-accuracy}
    \Description{{Time series plots showing absolute prediction errors (left Y) and normalized measured and predicted \CAC{} timeseries (right Y)  for \llamainf{}, LLaMA2 training, and Stable Diffusion (sdxl), alongside a table of the NRMSE in \CAC{} predictions for each workload.}}
\end{figure}

To strictly limit \edp{} and reduce transient jitter, we developed a digital power estimator (DPE) using a hardware event-counter based power estimation mechanism. 
This is similar to existing counter-based runtime power estimators~\cite{Simmani-MICRO52, PPEP-CPU-MICRO14, apollo2021} that use regression models to tune weights from hardware event counters.
However, unlike these solutions that use runtime power estimates for DVFS algorithms~\cite{Simmani-MICRO52, PPEP-CPU-MICRO14, apollo2021}, we tune the counter weights for $EDP_p$ management to have high accuracy in \textbf{medium to high} $C_{AC}$ scenarios, which is likely to cause \edp{}.
To do this, we use existing hardware counters that measure switching activity for different GPU pipelines, such as the matrix cores, vector and scalar ALUs, and caches.
The DPE performs a regression to tune the weights of these hardware event-counters to estimate $C_{AC}$, and therefore dynamic power.
Figure~\ref{fig:cac-timeseries} shows the correlation between our DPE and measured data when running ML applications, and Figure~\ref{fig:cac-nrmse} summarizes the Normalized Root Mean Squared Error (NRMSE) for each application.
These results show how generalizable the DPE is across a variety of ML workloads, not just inference.
The digital estimations and actual power track each other very well, with estimation error of 12\% or less (7\% geomean) across workloads.

Using this DPE, we developed a boost management algorithm that monitors power at sub-microsecond granularities and utilizes the AMD GPU's adaptive clock response mechanism~\cite{AMD-droop-variability} to throttle power.
At runtime, this mechanism estimates the total power consumption of the GPU using the DPE and compares it to a configurable threshold (e.g., 1.2$\times$TDP).
If the power estimate exceeds the threshold, the algorithm triggers the adaptive clock response to reduce the GPU's graphics clock frequency~\cite{AMD-droop-variability} and reduce power. 
Our experimental prototype can respond to \edp{} events at microsecond timescales of the DPE.
Section~\ref{subsec:eval-hw} shows results and insights from using this boost management mechanism on various inference models. 

\subsection{\texorpdfstring{\edp{}}{EDP} Management in Firmware}
\label{sec:design-fw}

The DPM varies the voltage and frequency of operation dynamically at a 1-2 ms cadence, which leads to two variability-inducing effects: (1) misalignment with application behavior since kernels (and therefore \edp{}) are much shorter than the firmware's operating granularity, and (2) firmware DVFS is not synchronized across GPUs, so when ML workloads run across multiple GPUs, PM firmware on different GPUs can set different frequencies for graphics and memory clocks at the same time. 
Thus, firmware frequency modulation can cause transient jitter and degrade overall application performance. 
Therefore, we examine whether controlling graphics and memory clock frequency limits allows PM firmware to operate deterministically, reduce kernel jitter, and improve performance.

For firmware \edp{} management, we identified two control parameters: graphics \textbf{frequency cap} ($f_{cap}$) and \textbf{memory clock state}. 
A frequency cap is an upper bound on the graphics clock frequency that constrains the PM firmware's DVFS range. 
Instead of allowing the firmware to boost freely up to the GPU's maximum supported frequency and get throttled due to \edp{}, the cap limits it to a specified frequency, reducing the range of frequency and indirectly limiting dynamic power and \edp{} variation. 
Memory clock states are discrete operating points for the GPU's memory subsystem, including the HBM clock and the memory controller/interconnect clock. 
Each state corresponds to a set of frequencies for these memory clocks.
PM firmware dynamically switches between these states based on memory bandwidth demand as well as the power consumed by competing compute resources. 
In addition to setting a graphics frequency cap, we identify levels at which to fix the memory clock state so that we avoid transient variation in memory clocks affecting power, \edp{}, and performance. 

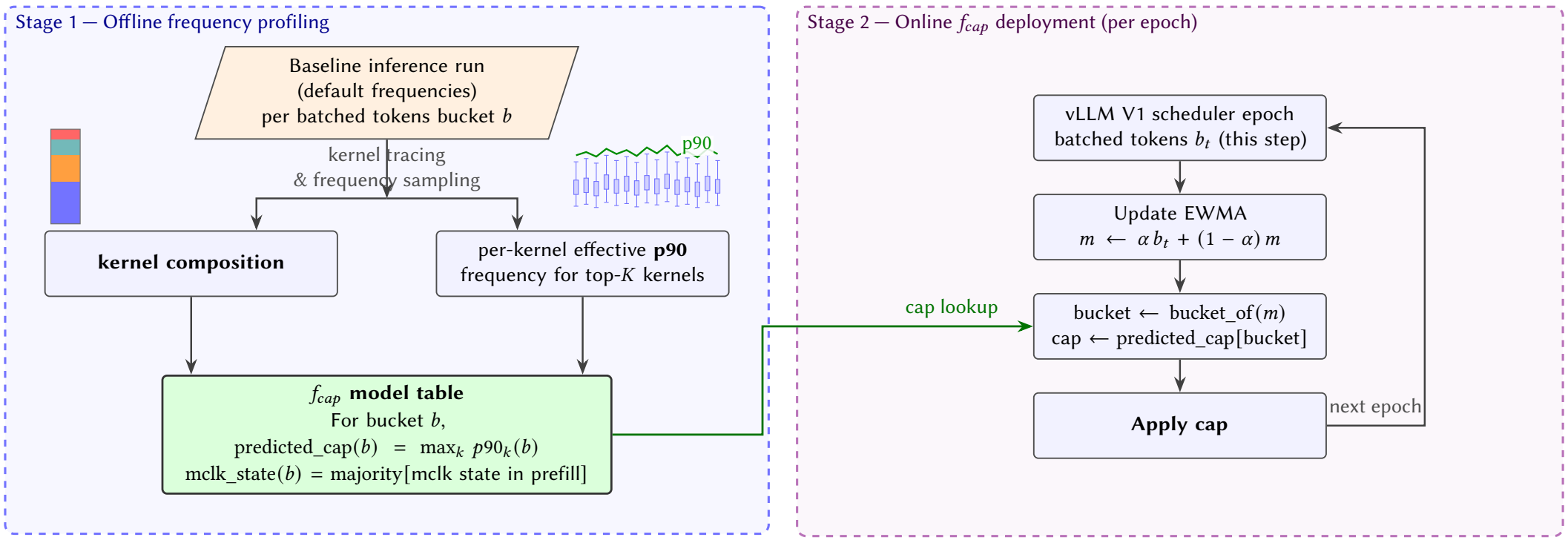
\begin{figure}[tb!]
  \centering
  \resizebox{\textwidth}{!}{%
\begin{tikzpicture}[
  node distance=5mm and 10mm,
  >={Stealth[length=2.4mm]},
  proc/.style={rounded corners=2pt, draw=black!70, fill=blue!5, line width=0.6pt,
               align=center, inner sep=4pt, text width=42mm, minimum height=10mm},
  data/.style={draw=black!70, fill=orange!12, line width=0.6pt, align=center,
               inner sep=5pt, text width=46mm, minimum height=10mm,
               trapezium, trapezium left angle=72, trapezium right angle=108,
               trapezium stretches body},
  model/.style={draw=black!80, fill=green!14, line width=0.9pt, align=center,
                rounded corners=2pt, inner sep=4pt, text width=66mm, minimum height=11mm},
  lbl/.style={text=black!70},
  flow/.style={->, line width=0.8pt, draw=black!75},
]

\node[data] (base) {Baseline inference run\\(default frequencies)\\per batched tokens bucket $b$};

\coordinate (split) at ($(base.south)+(0,-9mm)$);

\node[proc] (comp) at ($(base.south)+(-30mm,-19mm)$) {\textbf{kernel composition}};
\node[proc] (p90)  at ($(base.south)+( 30mm,-19mm)$)
  {per-kernel effective \textbf{p90}\\frequency for top-$K$ kernels};

\begin{scope}[shift={($(comp.north west)+(1mm,1mm)$)}]
  \def\bw{4.5mm}
  \fill[blue!55]   (0,0)      rectangle (\bw,6.5mm);
  \fill[orange!75] (0,6.5mm)  rectangle (\bw,10.5mm);
  \fill[teal!55]   (0,10.5mm) rectangle (\bw,13mm);
  \fill[red!60]    (0,13mm)   rectangle (\bw,14.5mm);
  \draw[black!55, line width=0.4pt] (0,0) rectangle (\bw,14.5mm);
\end{scope}
\node[inner sep=0pt] (barTop) at ($(comp.north west)+(3.25mm,17mm)$) {};

\begin{scope}[shift={($(p90.north east)+(-25mm,1.5mm)$)}]
  \def\hw{0.35} 
  \foreach \i/\low/\qone/\qthree/\high in {
    0/2.2/4.0/6.2/9.0,  1/2.5/4.3/6.5/9.5,  2/2.0/3.8/6.0/8.8,
    3/2.8/4.8/7.0/10.0, 4/2.3/4.2/6.3/9.2,  5/2.6/4.5/6.8/9.8,
    6/2.1/3.9/6.1/8.9,  7/2.7/4.7/6.9/10.2, 8/2.4/4.1/6.4/9.3,
    9/2.9/4.9/7.1/10.5, 10/2.2/4.0/6.2/9.0, 11/2.5/4.4/6.6/9.6,
    12/2.0/3.7/5.9/8.7, 13/2.8/4.6/6.8/10.0, 14/2.3/4.2/6.3/9.2
  } {
    \pgfmathsetmacro{\x}{1.475 + \i * 1.55}
    \pgfmathsetmacro{\xl}{\x-\hw}
    \pgfmathsetmacro{\xr}{\x+\hw}
    \draw[blue!55, line width=0.35pt] (\x mm, \low mm) -- (\x mm, \high mm);
    \draw[blue!55, line width=0.35pt] (\xl mm, \low mm) -- (\xr mm, \low mm);
    \draw[blue!55, line width=0.35pt] (\xl mm, \high mm) -- (\xr mm, \high mm);
    \fill[blue!18, draw=blue!55, line width=0.35pt]
      (\xl mm, \qone mm) rectangle (\xr mm, \qthree mm);
  }
  \draw[green!55!black, line width=0.75pt] plot coordinates {
    (1.475mm,10.0mm) (3.025mm,10.5mm) (4.575mm,9.8mm) (6.125mm,11.0mm)
    (7.675mm,10.2mm) (9.225mm,10.8mm) (10.775mm,9.9mm) (12.325mm,11.2mm)
    (13.875mm,10.3mm) (15.425mm,11.5mm) (16.975mm,10.0mm) (18.525mm,10.6mm)
    (20.075mm,9.7mm) (21.625mm,11.0mm) (23.175mm,10.2mm) };
  \node[text=green!55!black, inner sep=0.5pt,
        fill=white, fill opacity=0.85, text opacity=1]
    at (20mm,11.5mm) {p90};
\end{scope}
\node[inner sep=0pt] (freqTop) at ($(p90.north east)+(-12.5mm,18mm)$) {};

\node[model, below=36mm of base] (table)
  {\textbf{$f_{cap}$ model table}\\[1pt]
   For bucket $b$,\\[0.5pt]
   $\mathrm{predicted\_cap}(b)=\max_{k} \, p90_k(b)$\\[0.5pt]
   $\mathrm{mclk\_state}(b)=\mathrm{majority}[\text{mclk state in prefill}]$};

\draw[flow] (base.south) --
   node[lbl, align=center, inner sep=1.5pt]
        {kernel tracing\\\& frequency sampling} (split);
\draw[flow] (split) -| ([xshift=10mm]comp.north);
\draw[flow] (split) -| ([xshift=-10mm]p90.north);
\draw[flow] (comp.south) -- (comp.south |- table.north);
\draw[flow] (p90.south)  -- (p90.south  |- table.north);

\begin{scope}[on background layer]
  \node[draw=blue!55, dashed, rounded corners=4pt, line width=1pt,
        fill=blue!3, fit=(base)(comp)(p90)(barTop)(freqTop)(table),
        inner sep=6mm] (s1box) {};
  \coordinate (s2nw) at ([xshift=4mm]s1box.north east);
  \path let \p1=(s1box.east), \p2=(s1box.west),
            \p3=(s1box.north), \p4=(s1box.south),
            \n1={\x1-\x2}, \n2={\y3-\y4} in
    node[draw=violet!55, dashed, rounded corners=4pt, line width=1pt,
         fill=violet!3, minimum width=\n1, minimum height=\n2,
         inner sep=0pt, anchor=north west] (s2box) at (s2nw) {};
\end{scope}

\coordinate (s2hcenter) at ($(s2box.west)!0.5!(s2box.east)$);
\node[proc] (apply) at ($(s2hcenter)!0.58!(s2box.south)$) {\textbf{Apply cap}};
\node[proc, above=of apply] (bucket)
  {bucket $\leftarrow \mathrm{bucket\_of}(m)$\\cap $\leftarrow \mathrm{predicted\_cap}[\text{bucket}]$};
\node[proc, above=of bucket] (ewma)
  {Update EWMA\\$m \leftarrow \alpha\,b_t + (1-\alpha)\,m$};
\node[proc, above=of ewma] (epoch)
  {vLLM V1 scheduler epoch\\batched tokens $b_t$ (this step)};

\draw[flow] (epoch)  -- (ewma);
\draw[flow] (ewma)   -- (bucket);
\draw[flow] (bucket) -- (apply);
\draw[flow] (apply.east) -- ++(15mm,0)
   node[lbl, pos=0.5, above]{next epoch} |- (epoch.east);

\draw[flow, draw=green!45!black, line width=0.9pt]
   (table.east) -- ++(23mm,0) |-
   node[lbl, pos=0.85, above, text=green!45!black]{cap lookup} (bucket.west);

\node[anchor=north west, text=blue!55!black] at (s1box.north west)
  {\,Stage 1 --- Offline frequency profiling};
\node[anchor=north west, text=violet!55!black] at (s2box.north west)
  {\,Stage 2 --- Online $f_{cap}$ deployment (per epoch)};

\end{tikzpicture}%
  }
  \caption{\SolutionName{}'s two stages. Offline, a baseline inference run is traced
  per batched-token bucket to recover kernel composition and per-kernel p90
  effective frequencies, which together yield an $f_{cap}$ model table. Online,
  the vLLM scheduler's batched-token count is mapped to a
  bucket, and used to look up and apply the corresponding cap per epoch.}
  \Description{\SolutionName{}'s two stages. Offline, a baseline inference run is traced
  per batched-token bucket to recover kernel composition and per-kernel p90
  effective frequencies, which together yield an $f_{cap}$ model table. Online,
  the vLLM scheduler's batched-token count is mapped to a
  bucket, and used to look up and apply the corresponding cap per epoch.}
  \label{fig:solution-method}
\end{figure}

We developed a two-stage algorithm, called \textbf{\SolutionName{}}, for selecting optimal values for these control parameters when serving inference requests online, since the optimal frequencies are workload- and input-dependent. 
We categorize possible input lengths (number of batched tokens) for LLM inference prompts into buckets $b$, up to the maximum number of batched tokens for each model. 
We can then determine an optimal graphics frequency cap $f_{\mathrm{cap}}$ as well as fixed memory clock state for each batched-token length bucket $b$ based on offline frequency profiling.
This allows us to construct a lookup table for optimal $f_{cap}$s and memory clock states which can be used online without needing to re-profile incoming requests.  
Thus, \SolutionName{} involves two stages: (1) developing an offline $f_{\textrm{cap}}$ model for graphics and memory clock states, and (2) deploying the frequency states dynamically based on input request characteristics
(Figure~\ref{fig:solution-method}).

\subsubsection{Stage 1: Offline frequency profiling}
\label{subsubsec:stage1}
For every batched-token bucket $b$, we run baseline inference at the GPU's default (uncapped) frequency and collect kernel traces together with per-kernel effective frequency samples.
From this profiling, we obtain the \emph{kernel composition} of each transformer layer, showing which kernels dominate overall prefill time.
We use this kernel composition to extract the top-$K$ kernels in the workload ($K$ is a per-model, heuristic value -- we select $K$ such that the kernels together constitute $>50\%$ of the workload's runtime).
Section~\ref{subsec:eval-hw} shows us that for frequency capping to not degrade performance, we must allow compute-intensive kernels to computationally sprint to completion. 
Therefore, from the profiled frequency samples for each of the top-$K$ kernels, we compute the 90th-percentile frequency $p90_k(b)$. 
Because a single kernel that still needs high frequency would otherwise become the latency bottleneck (degrading performance), the predicted cap for the bucket is set by the most demanding of those kernels:
\begin{equation}
  \mathrm{predicted\_cap}(b)
    \;=\;
    \max_{k \in \mathcal{K}(b)} p90_k(b),
  \label{eq:fcap}
\end{equation}
where $\mathcal{K}(b)$ is the top-$K$ kernel set for bucket $b$.
Alongside the frequency cap, we also record the memory clock state for each bucket, setting $\mathrm{mclk\_state}(b)$ to the memory clocks that the GPUs spend the majority of their time in during the profiled baseline run.
The result is a static lookup table $\{b \mapsto (f_{\mathrm{cap}}(b),\,\mathrm{mclk\_state}(b))\}$ that is later consumed at serving time.


\subsubsection{Stage 2: Online $f_{\mathrm{cap}}$ deployment.}
At serving time the table is applied once per scheduler epoch. Let $b_t$ be the
number of batched tokens scheduled in epoch $t$. 

We track a smoothed batch size with an exponentially weighted moving average to bias the input towards the latest epoch while avoiding firmware oscillation on short-lived batch-size spikes: 
\begin{equation}
  m_t
    \;=\;
    \alpha\, b_t + (1-\alpha)\, m_{t-1},
  \qquad 0 < \alpha \le 1,
  \label{eq:ewma}
\end{equation}
We use $\alpha = 0.3$ for all our experiments, although $\alpha$ can be tuned depending on how reactive the system should be to abrupt request length changes. 

We then map $m_t$ onto a bucket, and look up the corresponding cap and memory clock state:
\[
  \mathrm{bucket} \leftarrow \mathrm{bucket\_of}(m_t),
  \\
  f_{\mathrm{cap}} \leftarrow \mathrm{predicted\_cap}[\mathrm{bucket}],
  \\
  \mathrm{mclk\_state} \leftarrow \mathrm{mclk\_state}[\mathrm{bucket}].
\]
The cap is then applied to the GPUs, and this process repeats whenever a new request arrives.

\section{Methodology}
\label{sec:methodology}

\subsection{Serving stack and hardware}
All models are served with vLLM~\cite{kwon2023efficient} running inside a \texttt{rocm/vllm:latest} docker container (ROCm 7.0.0, PyTorch 2.9.0a0, Ubuntu 22.04.5) on a single node of 8x\threeX{} GPUs~\cite{cdna3}.
We built an experimental system on top of a production quality \threeX. 
Table~\ref{tab:inf-workloads} provides details for both offline and online inference models and configurations we ran in Section~\ref{sec:eval}.
We ran offline inference on a single GPU across dense and MoE models, and across different batch sizes and input lengths. 
For online serving, we ran with tensor parallelism TP=8. 
The graphics frequency on MI300X can be varied up to 2100 MHz. 
Memory clock states can vary between 3 states corresponding to clock frequencies of 1300 MHz, 1550 MHz, and 1800 MHz. 
We drive load with vLLM's \texttt{bench serve} harness, replaying a sampled set of 500 LongBench~\cite{bai2025longbenchv2deeperunderstanding} prompts under a fixed seed so all runs have an identical request stream.
For DeepSeek-V3.2, the maximum number of batched tokens is limited to 8k since we run out of memory (OOM) for weights and activations beyond this limit. 
While larger prompts work, vLLM will divide them into 8k chunks (chunked prefill), so analyzing up to 8k prompts is sufficient. 
Since \edp{} primarily affects prefill behavior (Section~\ref{sec:exbehavior}), we run in a prefill-decode disaggregated scenario~\cite{patel-2024-splitwise, Zhong-DistServe-OSDI24} and requests arrive as a Poisson process at different request rates. 

%
%
\begin{scriptsize}
\begin{table}[tb!]
\centering
\caption{Inference workloads. Offline inference ran on a
single MI300X GPU, while online serving is run tensor parallel across 8 \threeX{} GPUs
with Poisson arrival at varying QPS rates.}
\label{tab:inf-workloads}
\setlength{\tabcolsep}{4pt}
\begin{tabular}{@{}lllllll@{}}
    \toprule
    \textbf{Scenario} & \textbf{Model} & \textbf{Type} & \textbf{dtype} & \textbf{Dataset} & \textbf{Config} & \textbf{TTFT SLO} \\ \midrule
    \multirow{4}{*}{Offline}
      & Qwen3-14B~\cite{qwen3}                                        & Dense     & BF16 & --        & bsz 1, ip 4k, op 128            & N.A.    \\
      & LLaMA3-8B~\cite{touvron-2023-llama,dubey2024llama3herdmodels} & Dense     & BF16 & --        & bsz 1/4/8/16, ip 1k/4k, op 128  & N.A.    \\
      & LLaMA3-70B~\cite{touvron-2023-llama,dubey2024llama3herdmodels}& Dense     & BF16 & --        & bsz 1/4, ip 1k/4k, op 128       & N.A.    \\
      & Mixtral-8x7B~\cite{Mixtral}                                   & MoE       & BF16 & --        & bsz 1/4, ip 1k/4k, op 128       & N.A.    \\ \midrule
    \multirow{3}{*}{Online}
      & Llama-3.1-405B~\cite{dubey2024llama3herdmodels}               & Dense     & FP8       & LongBench & 500 req, ip 8,894 tok (avg)  & 4.62\,s \\
      & DeepSeek-V3.2                                                 & MoE (MLA) & FP8       & LongBench & 500 req, ip 4,528 tok (avg)  & 2.08\,s \\
      & Mistral-Medium-3.5-128B                                       & Dense     & FP8       & LongBench & 500 req, ip 11,112 tok (avg) & 2.08\,s \\
    \bottomrule
\end{tabular}
\vspace{-2ex}
\end{table}
\end{scriptsize}

\subsection{Request Rates}

We sweep request rate or queries per second (QPS) up to 0.3 because that is the point at which the prefill pipeline approaches saturation for the long-context prompts from LongBench. 
Treating prefill as a queue, the offered load (utilization $\rho$) is the ratio of the arrival token rate to the sustained prefill token throughput of the MI300X node. 
Arrival token rate is QPS times the LongBench average input length (roughly 4.5k–11k tokens per request depending on model), while prefill token throughput sustained by the MI300X node is on the order of $\sim$3.4–3.8k tokens/s.
Under this model, utilization grows roughly linearly with QPS and reaches $\rho \approx 1$ near 0.35 QPS.
Therefore, request rates beyond 0.3 QPS drive the system into an unstable, backlog-dominated regime that no longer reflects steady-state serving, which is why we bound the sweep at 0.3.

\subsection{Metrics}

We perform measurements for the prefill portion of the run since it suffers the most from \edp{} (Section~\ref{sec:exbehavior}).
For single GPU offline inference, we show percentage improvements in Time-To-First-Token (TTFT) and reduction in \edp{}.
We also show normalized kernel durations to highlight the effect of \edp{} reduction on transient variability. 
For online serving, we report TTFT at three percentiles to account for bursty Poisson arrival of requests: the median (p50), which reflects typical user-facing latency; 90th percentile (p90), which captures latency for the slower tail of requests; and 99th percentile (p99), which reflects worst-case behavior.
We also report \%\emph{SLO attainment}, defined as the percentage of requests whose TTFT falls at or below an SLO threshold $\tau$. 
Following prior serving work~\cite{sola,alpaserve,dynamollm}, we set $\tau = k \cdot T_{\text{mean}}$.
$T_{\text{mean}}$ is the model's mean TTFT when serving a single LongBench request on an idle TP=8 deployment, and $k$ is a slowdown budget, set to 2 based on recommendation from prior work~\cite{sola,tediserve} for long context workloads, since LongBench prompts average 4.5k--11k tokens for the models we serve.
Table~\ref{tab:inf-workloads} lists the resulting per-model $\tau$.

\section{Evaluation}
\label{sec:eval}

\subsection{\texorpdfstring{\edp{}}{EDP} Management in Hardware}
\label{subsec:eval-hw}
Using the DPE and hardware boost management mechanism outlined in Section~\ref{sec:design-hw}, we determine the impact of \edp{} clamping by varying the acceptable \edp{} threshold from 
1.0$\times$ to 1.5$\times$TDP. 
We first show results for \llamainf{} (4096 input tokens, 128 output tokens, and batch size 1) where the DPE estimation error is about 6\%.
Then we show results for \llamamodel{} in different input configurations, as well as \qwenmodel{}, \mixtralmodel{}, and LLaMA3-70B (Table~\ref{tab:bmks}).

\begin{figure}[tb!]
    \centering
    \begin{subfigure}{0.4\linewidth}
        \centering
        \includegraphics[width=\linewidth]{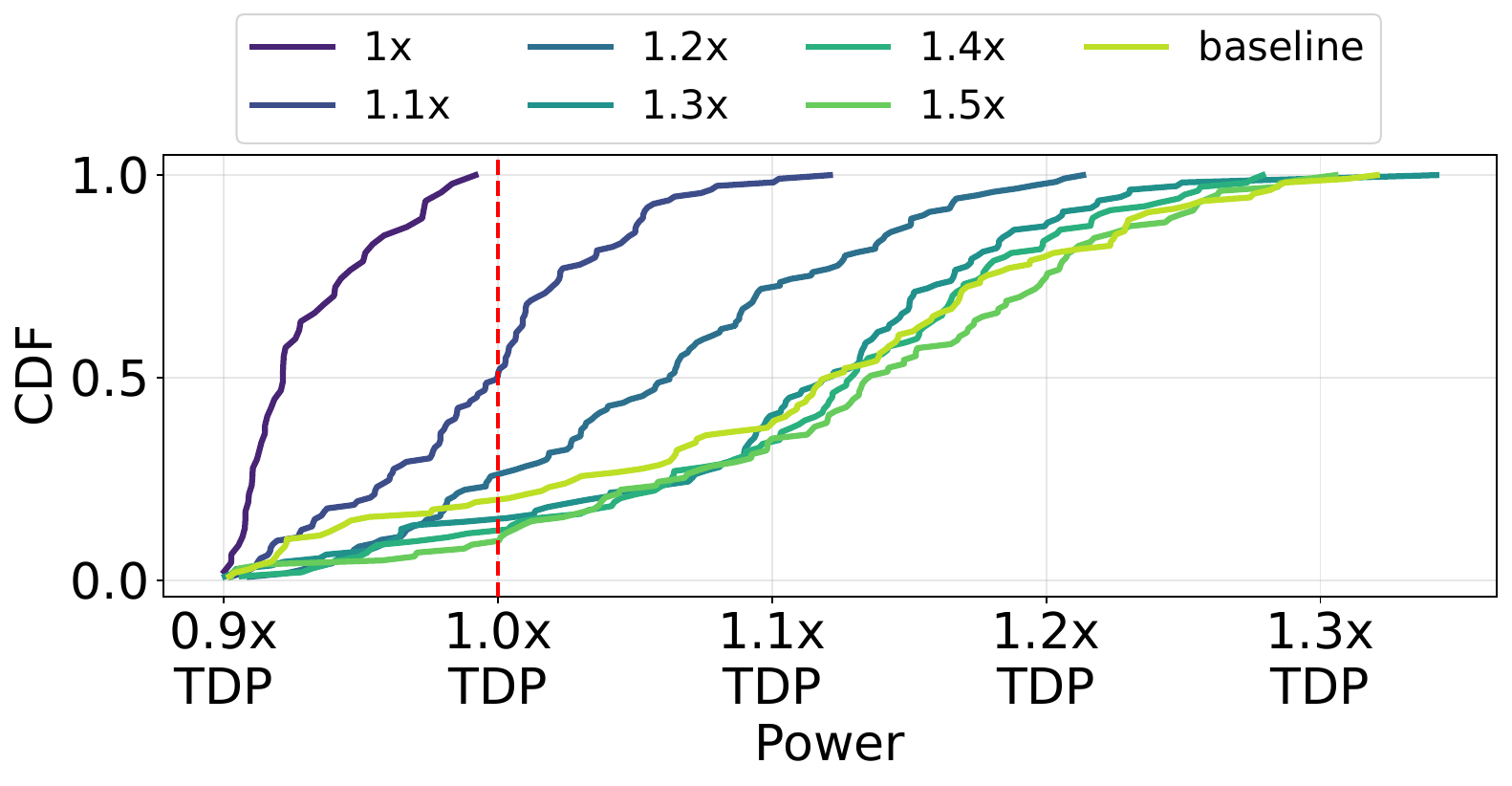}
        \caption{}
        \label{subfig:edc-cdf}
    \end{subfigure}
    \hfill
    \begin{subfigure}{0.5\linewidth}
     \centering
        \includegraphics[width=\linewidth]{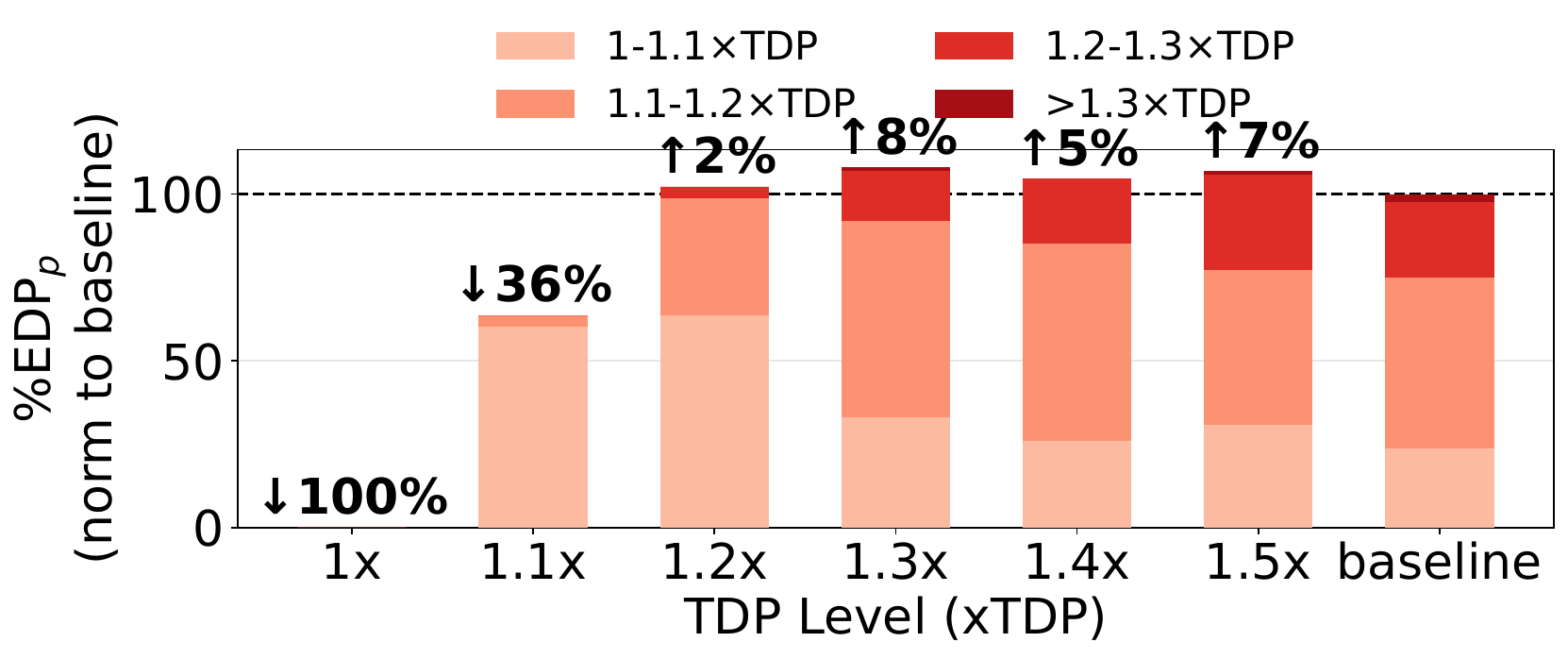}
        \caption{}
        \label{subfig:edc-edp-dist}        
    \end{subfigure}
    \hfill
    \begin{subfigure}{0.45\linewidth} 
        \centering
        \includegraphics[width=\linewidth]{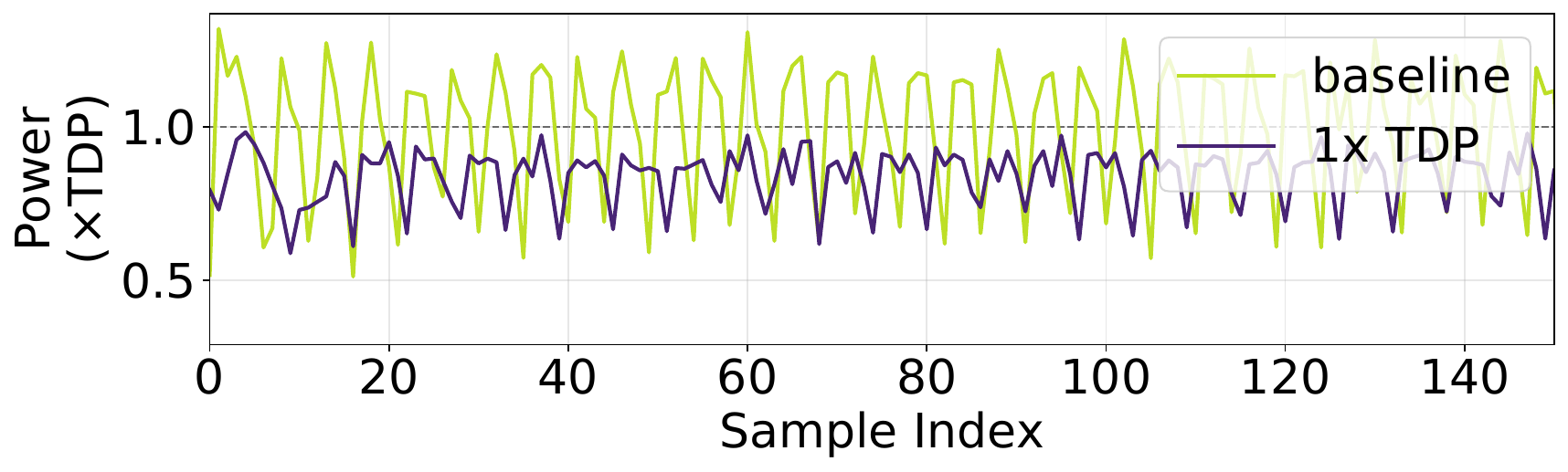}
        \caption{}
        \label{subfig:edc-deltapower}        
    \end{subfigure}
    \hfill    
    \begin{subfigure}{0.45\linewidth} 
        \centering
     \includegraphics[width=\linewidth]{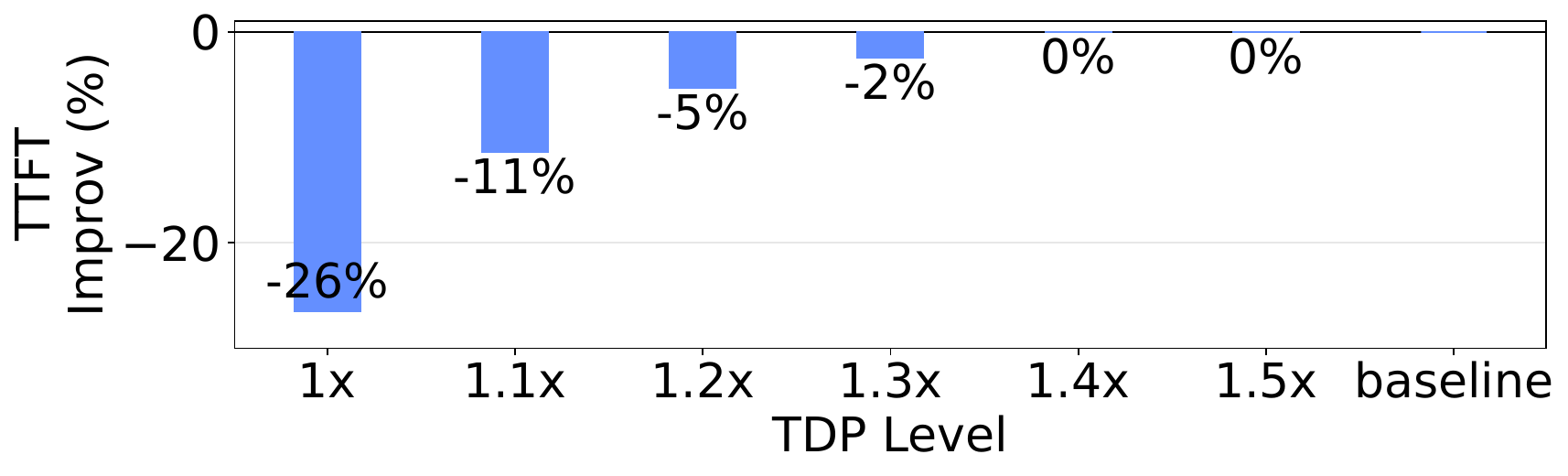}
        \caption{}
        \label{subfig:edc-clamp-ttft}
    \end{subfigure}
    \vspace{-2ex}
    \caption{(a) Cumulative distribution of \edp{} when boost management clamps power to different TDP levels; (b) Reduction in \%\edp{} at different clamping levels; (c) Timeseries showing varying power values for \llamainf{} with and without clamping; and (d) Performance impact relative to the unclamped baseline case (\%). All results are for the prefill portion of \llamainf{} with 4096 input tokens and a batch size of 1.}
    \Description{(a) Cumulative distribution of \edp{} when the boost management mechanism clamps power to different TDP levels; (b) Reduction in \%\edp{} at different clamping levels, (c) Timeseries showing varying power values for \llamainf{} with and without clamping and (d) Performance impact relative to the unclamped baseline case (\%). All results are for the prefill portion of \llamainf{} bs1 ip4096.}
    \label{fig:clamp-results}
    \vspace{-2ex}
\end{figure}

Figure~\ref{subfig:edc-cdf} shows the power measurement CDF for \llamainf{}. 
We utilize the DPE to cap the boost from 1$\times$TDP to 1.5$\times$TDP and compare it to the baseline normal operation without clamping (\Edit{labeled \textit{baseline}}).
First, we observed that the boost management mechanism effectively clamps power based on the set threshold.
The CDF of power events for each threshold shows that very few ($<$2\%) events extend beyond the specified threshold value.
Moreover, \textit{baseline}'s \edp{} behavior has higher amplitude power overshoots than those managed by our boost management mechanism.
The lower boost thresholds (up to 1.1$\times$TDP) have significantly fewer overshoots and fewer higher magnitude overshoots than \textit{baseline}.
At 1.2$\times$TDP and above, the numbers of overshoots are similar to the baseline, but on average they have lower peak amplitudes.

Figure~\ref{subfig:edc-deltapower} shows the timeseries of the power for the 32 Transformer layers in \llamainf{} for prefill for the unclamped and fully clamped (1$\times$TDP) case.
Power varies across each Transformer layer in the unclamped case either due to application behavior and/or \edp{} mitigation.
However, by using the boost cap, the fully clamped solution (1$\times$TDP) removes these power peaks.
Unfortunately, Figure~\ref{subfig:edc-clamp-ttft} demonstrates that the performance impact of eliminating power peaks (by limiting boost to 1.0$\times$TDP) is severe.
The TTFT for inference degrades by 26\% compared to normal operation when power is clamped to 750W (TDP).
Conversely, adding more boost headroom improves performance until it matches the \textit{baseline} at 1.4$\times$TDP.
This shows that boosting has computational sprinting benefits, allowing workloads to sprint during \edp{} events. 

\begin{figure}[tb!]
    \centering
    \begin{subfigure}{\linewidth}
        \centering
        \includegraphics[width=0.8\linewidth]{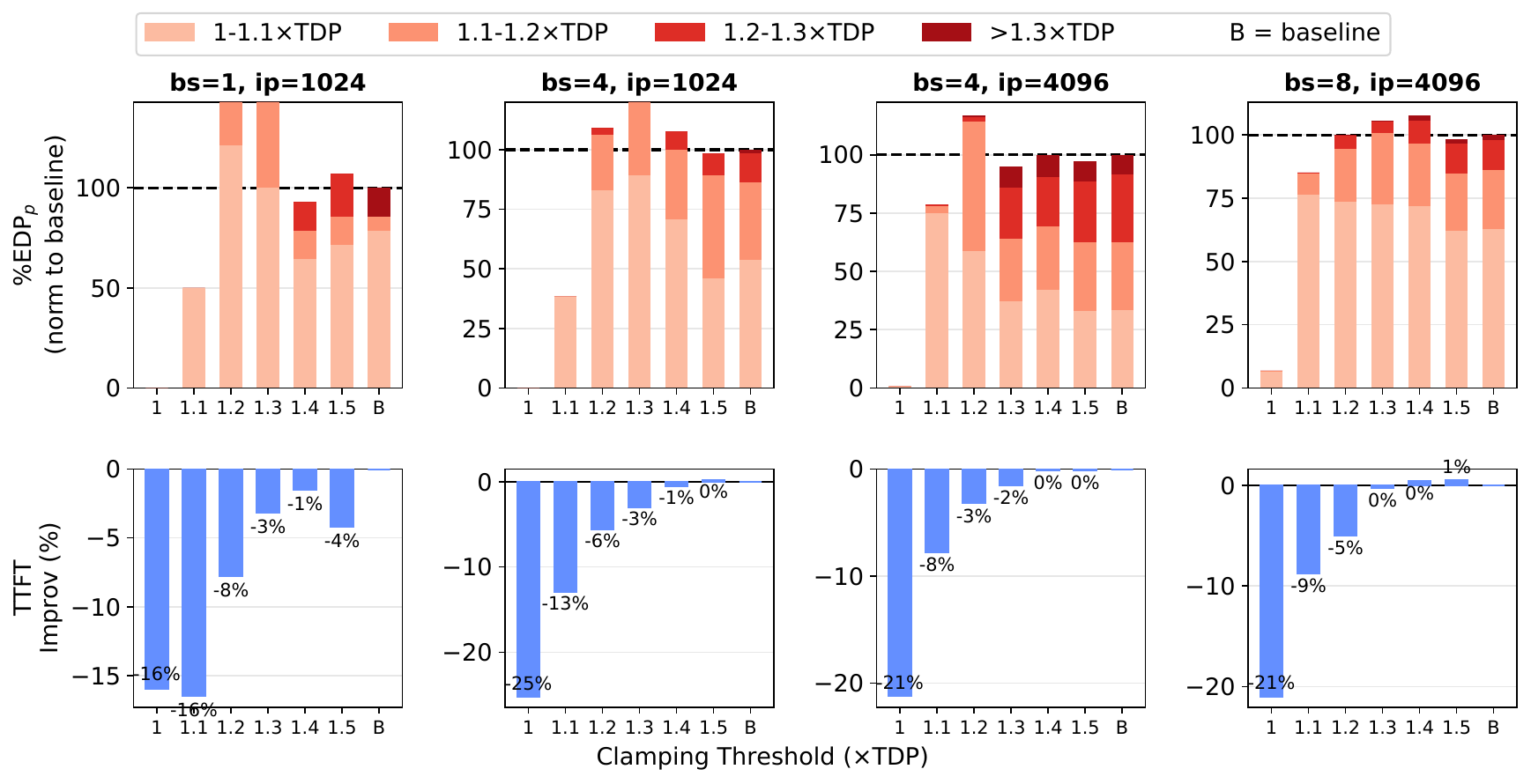}
        \caption{\Edit{\llamainf{} for different input lengths/batch sizes}}
        \label{subfig:clamp-all-configs}
    \end{subfigure}
    \begin{subfigure}{\linewidth}
     \centering
        \includegraphics[width=0.8\linewidth]{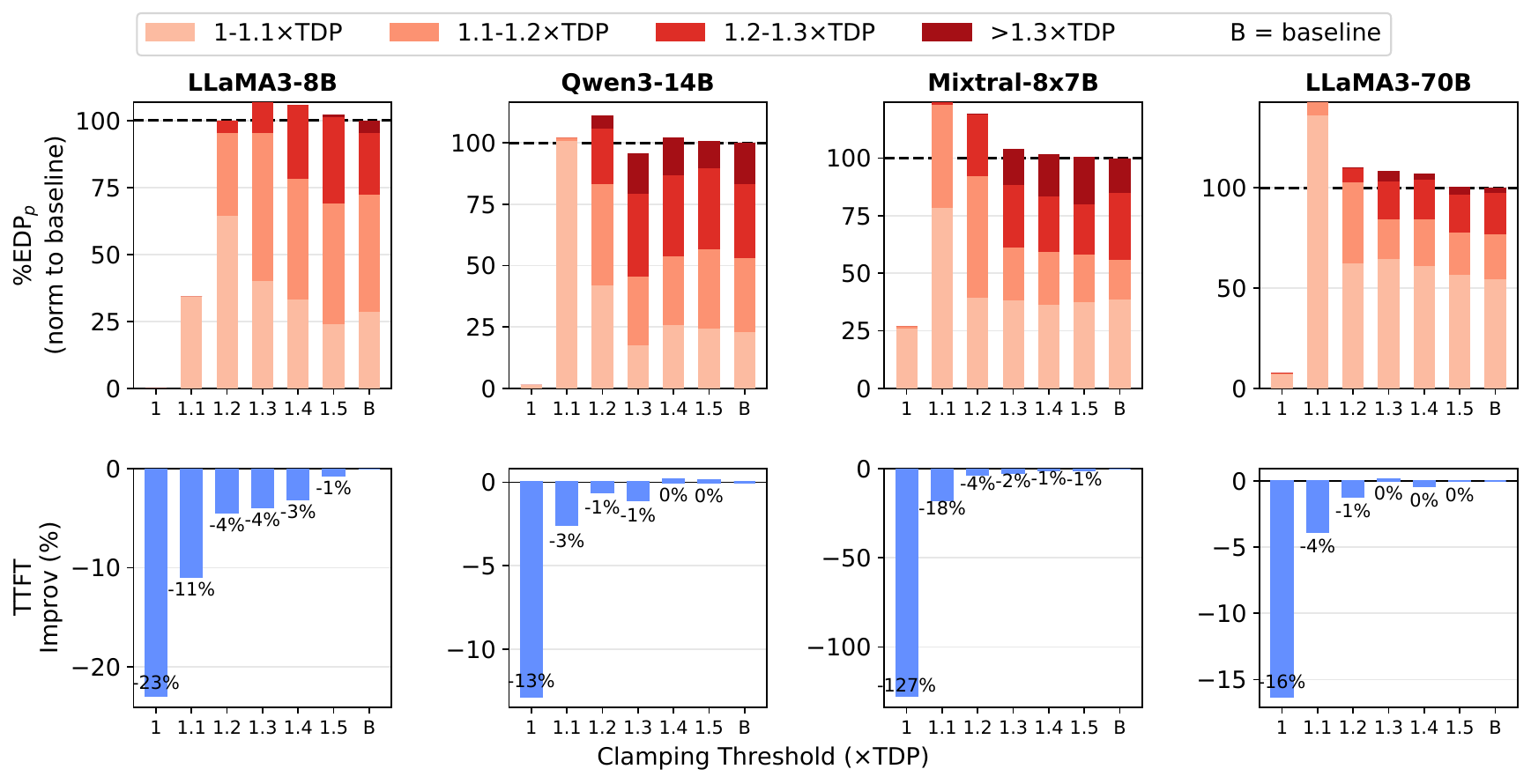}
        \caption{\Edit{bs1, ip4096 configuration for different models}}
        \label{subfig:clamp-all-models}        
    \end{subfigure}
    \caption{\%\edp{} and performance impact at different clamp levels for (a) varying batch size and input lengths for \llamainf{} and (b) different models with bsz 1 and 4096 input tokens.}
    \label{fig:clamping-all}
    \Description{Clamping results showing \%\edp{} at different clamping levels, and performance impact for (a) different batch size and input configurations for \llamainf{} and (b) different models with batch size 1 and 4096 input tokens}
\end{figure}

We also evaluate clamping across different models, input lengths, and batch sizes. 
Figure~\ref{subfig:clamp-all-configs} shows clamping effects for different batch sizes and input configurations for LLaMA3-8B inference.
Across different input configurations, \llamainf{} shows similar \edp{} and performance trends with increasing clamping thresholds.
As batch size increases, the proportion of high magnitude \edp{} events ($>1.2\times$TDP) reduces in the prefill phase of the unclamped baseline. 
However, lower clamping thresholds lose out on sprinting benefits and suffer from worse performance. 
Figure~\ref{subfig:clamp-all-models} shows other models exhibit the same trend (geomean 28\% loss at 1$\times$TDP clamp). 
The proportions of \edp{} in the unclamped baseline (labeled B) are different for different models, with \qwenmodel{} and \mixtralmodel{} showing more high magnitude \edp{} events than the LLaMA models.
\mixtralmodel{} in particular suffers severe performance degradation when the threshold is limited to 1.0$\times$TDP, reinforcing \edp{}'s sprinting benefits. 
Regardless, a clamping threshold of 1.2-1.3$\times$TDP is sufficient to eliminate high-magnitude \edp{} for all these models, while providing near-identical performance to the \textit{baseline}.
Thus, Figure~\ref{subfig:clamp-all-models} shows the DPE  effectively clamps \edp{} even for models it is not tuned on, like Qwen3-14B and Mixtral8x7B.

Overall, these results show that \edp{}'s computational sprinting is crucial for highly variable applications like LLMs.
However, we do not require a significant boost beyond the TDP.
Instead, some degree of \edp{} beyond TDP (e.g., 1.3$\times$TDP) is sufficient.

\takeawaybox{\begin{itemize}
    \item Hardware clamping mechanisms can be tuned to strictly control \edp{} to specific limits. 
    \item Low-magnitude \edp{} allows workloads to \textit{sprint} for short bursts of time by boosting their power. This boost leads to better performance for highly variable applications like LLMs.
    \item Sprinting benefits diminish beyond a certain degree (1.3$\times$TDP) -- higher magnitude \edp{} does not yield better performance over baseline. 
\end{itemize}
}

\subsection{\texorpdfstring{\edp{}}{EDP} Management in Firmware}
\label{subsec:eval-fw}

\subsubsection{Single GPU Offline Inference}
\label{subsubsec:single-gpu}
To determine how well existing DPM solutions deal with the variable nature of ML applications and their associated overshoot behavior, we ran three solutions with various frequency caps:

\noindent
\textbf{\textit{Oracle Capping}}: \textit{Oracle Capping} shows the performance for an idealized, unattainable DVFS solution.
It detects the best operating frequency for each kernel during the frequency cap sweep and analytically combines 
them with zero overhead.

\noindent
\textbf{\textit{Realistic Oracle}}: This approach includes the overhead of predicting and transitioning to the best frequency, while it still uses oracle frequencies.
Thus, it only modifies frequency approximately every two milliseconds -- the granularity of traditional firmware V/f scaling~\cite{BharadwajDMBK23, Park-DVFS-Modeling, GreenGoverners, DPM-CCGRID}.

\noindent
\textbf{\textit{E2E Capping}}: \textit{E2E Capping} 
uses the best operating frequency for each application when running with a maximum frequency cap.
ML applications often operate at a lower effective frequency due to their high compute demand and resulting power draw~\cite{greenllm, mlperf25}.
Thus, this approach determines if lower frequencies alleviate the negative impact of \edp{} management, reduce power throttling, and improve performance.
\textit{E2E Capping} is the mechanism used for \SolutionName{}.
We compare these to the oracle solutions to identify how close \SolutionName{} is to an idealized, unattainable solution.
We use the methodology in Section~\ref{subsubsec:stage1} to identify optimal frequency caps for each model and input.

\begin{figure}[tb!]
    \centering
    \includegraphics[width=0.7\linewidth]{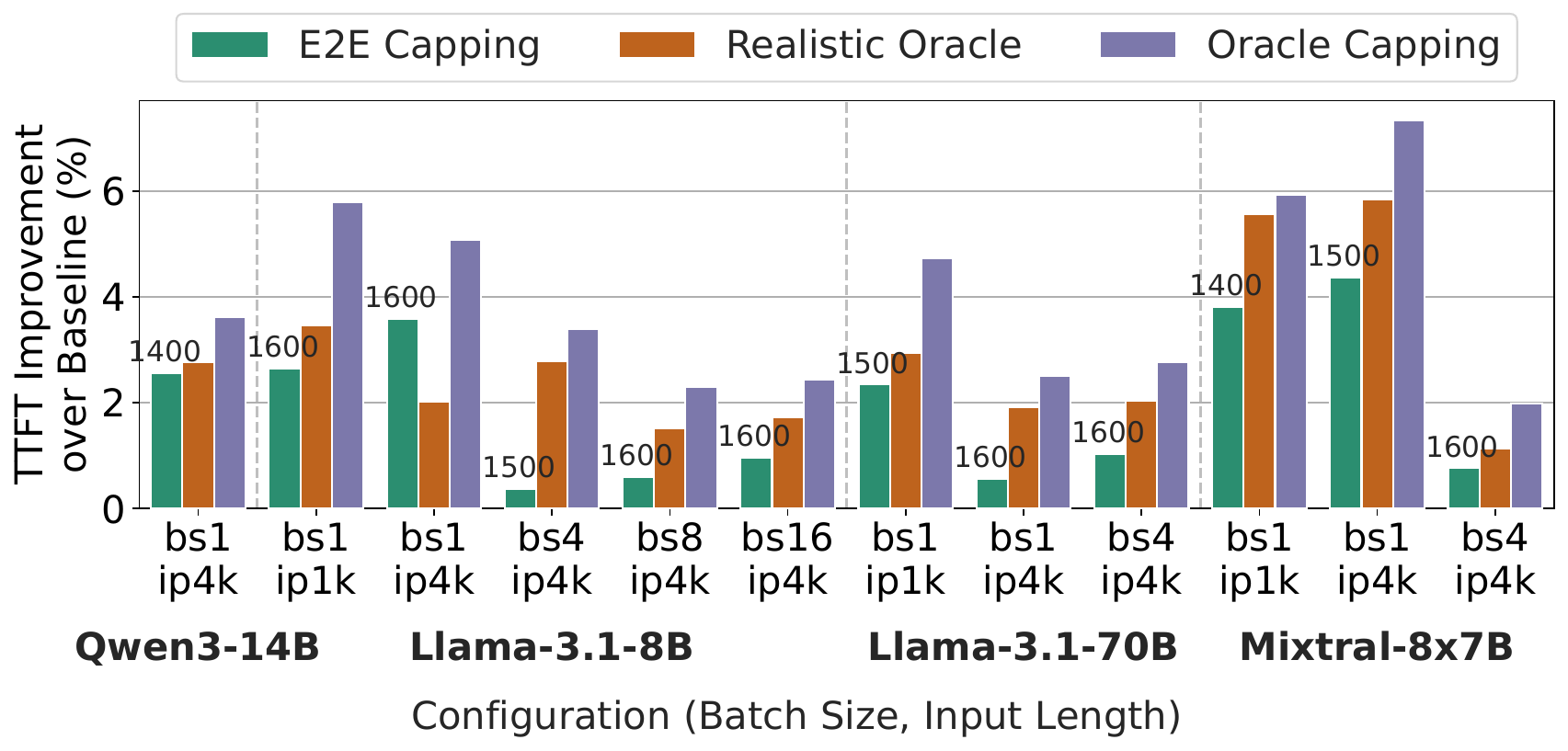}
    \vspace{-2ex}
    \caption{\%TTFT improvement over uncapped (2100 MHz $f_{max}$) baseline with \textit{E2E Capping} and \textit{oracle} solutions.}
    \label{fig:oracle-barplots}
    \Description{\%TTFT improvement over uncapped (2100 MHz maximum gfxclk frequency) baseline with \textit{E2E Capping} and \textit{oracle} solutions.}
    \vspace{-2ex}
\end{figure}

Figure~\ref{fig:oracle-barplots} shows how our three firmware 
solutions improve prefill performance relative to a default run with no frequency capping (\textit{Baseline}).
Although the power management algorithm does a reasonable job, there is still the potential for additional performance if we can identify the best operating frequency for each application.
\textit{Oracle Capping}'s results demonstrate that the existing power management algorithm is reactive and accurate enough to produce viable, effective DVFS results for each application. 
\textit{Oracle Capping} provides the most performance improvement ($\approx$7\%) for workloads like \mixtralmodel{}, which has very short kernels.
However, other applications with larger input sizes, and hence longer kernels, see diminishing performance benefits with \textit{Oracle Capping}. 
Although \textit{Realistic Oracle} cannot change quickly enough to capture all of \textit{Oracle Capping}'s performance benefits, it still improves performance by 3\% geomean (6\% max) over \textit{Baseline}.
Finally, \textit{E2E Capping} 
identifies 
a frequency cap that limits \edp{} events and enables the DPM to operate more effectively.
Thus, \textit{E2E Capping} improves performance by up to $\approx$4\% over \textit{Baseline} on real hardware, albeit not to the same extent as \textit{Oracle Capping}.
\begin{figure}[tb!]
    \centering
    \includegraphics[width=0.7\linewidth]{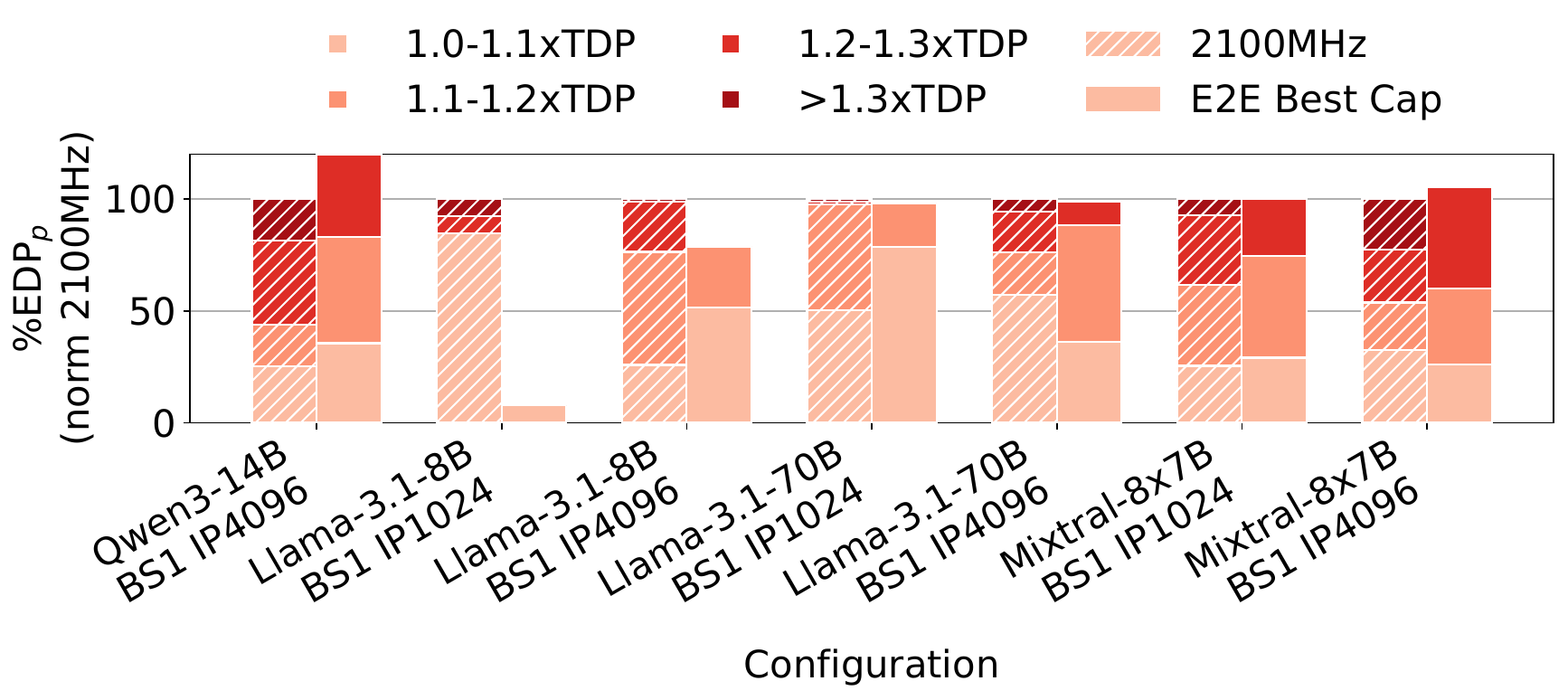}
    \vspace{-2ex}
    \caption{Reduction in high-magnitude \edp{} (normalized to 2100MHz run) with \textit{E2E Capping} for a subset of configurations from Figure~\ref{fig:oracle-barplots}.}
    \label{fig:delta-power-capping}
    \Description{Reduction in high-magnitude \edp{} (normalized to 2100MHz run) with \textit{E2E Capping} for a subset of configurations from Figure~\ref{fig:oracle-barplots}.}
\end{figure}

\begin{figure*}[tb!]
    \begin{subfigure}{0.32\linewidth}
        \includegraphics[width=\linewidth]{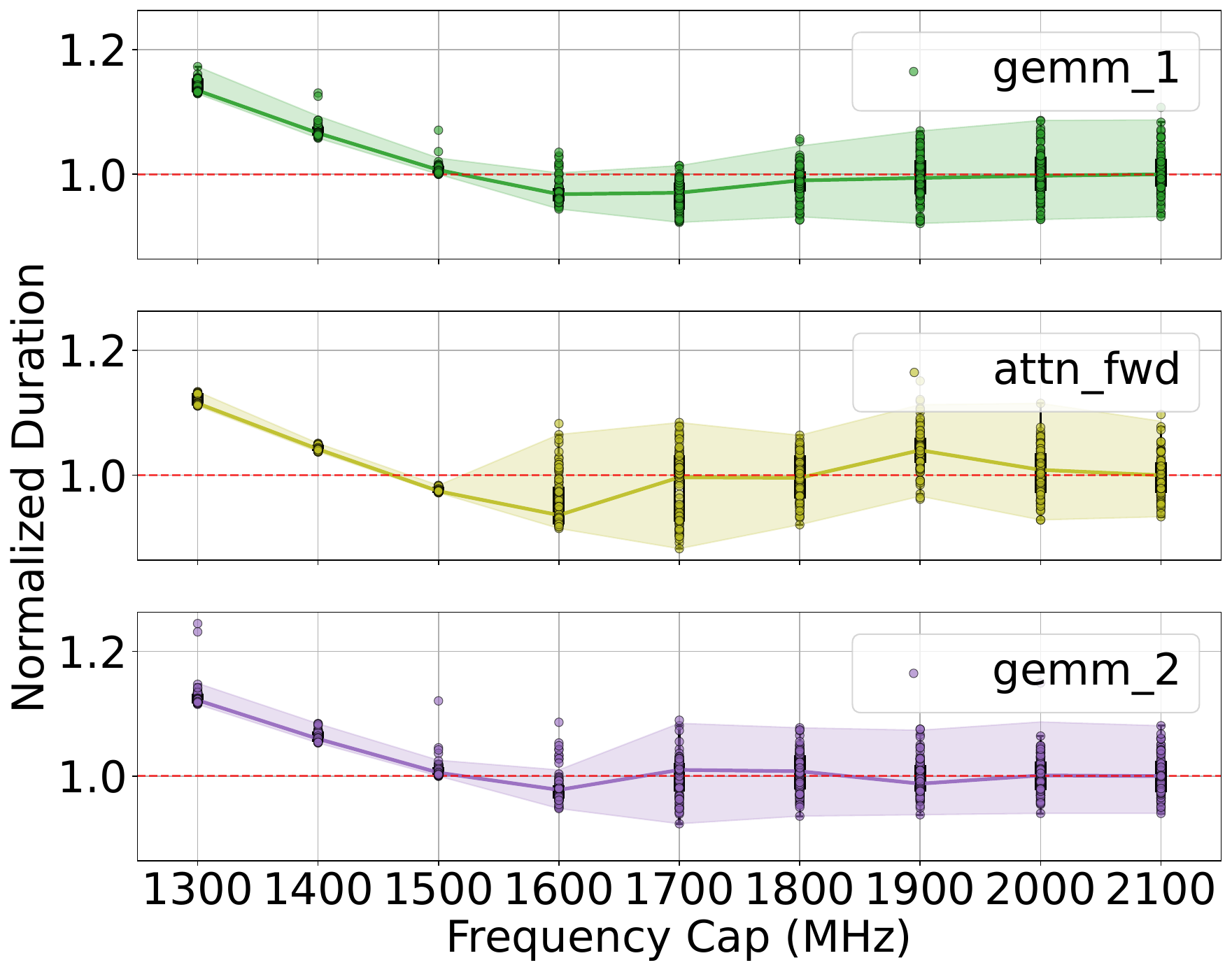}
        \caption{\llamamodel{} bs1}
        \label{subfig:llama3-8b}
    \end{subfigure}
    \hfill
    \begin{subfigure}{0.32\textwidth}
        \includegraphics[width=\linewidth]{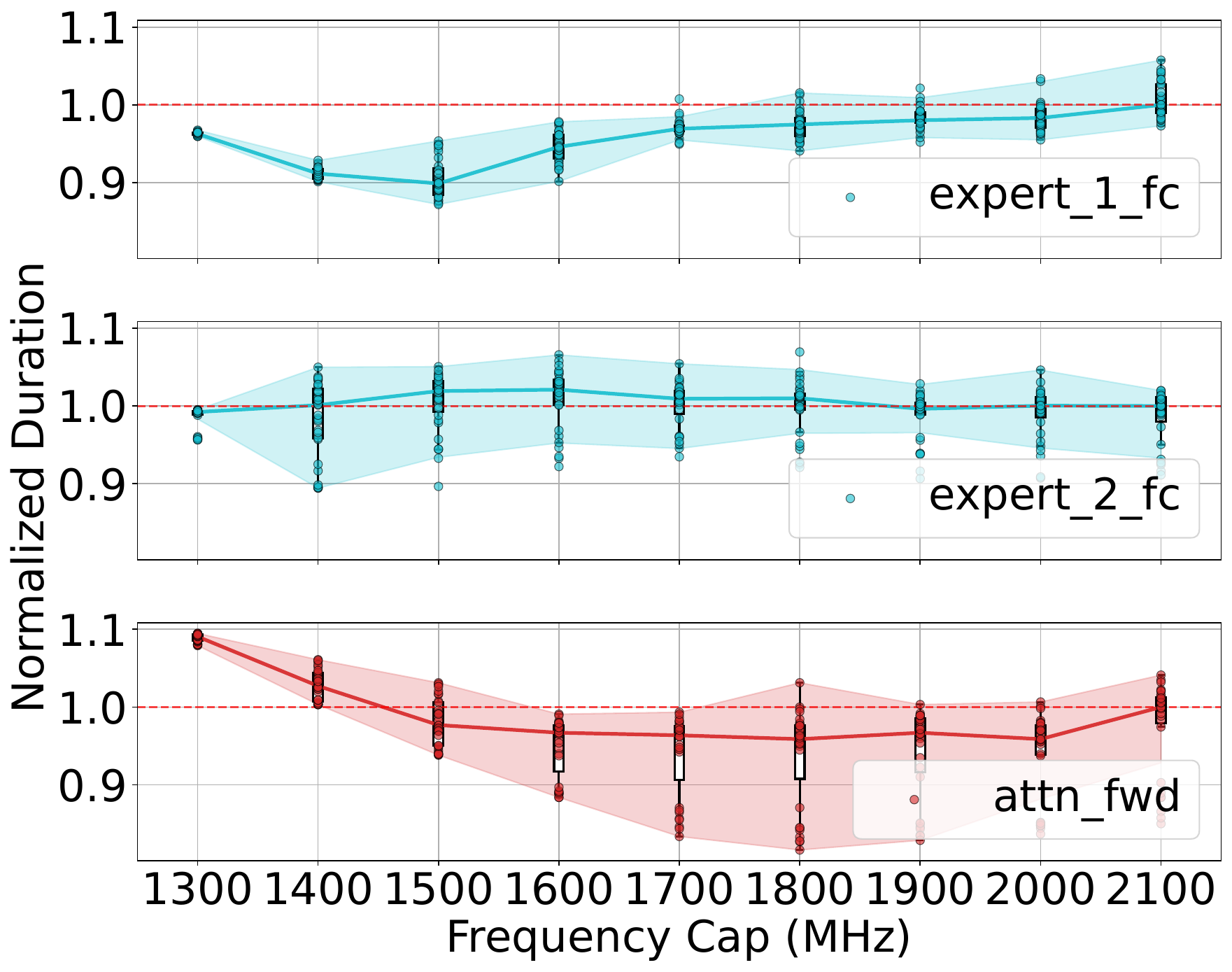}
        \caption{\mixtralmodel{} bs1}
        \label{subfig:mixtral}
    \end{subfigure}
    \hfill
    \begin{subfigure}{0.32\textwidth}
        \includegraphics[width=\linewidth]{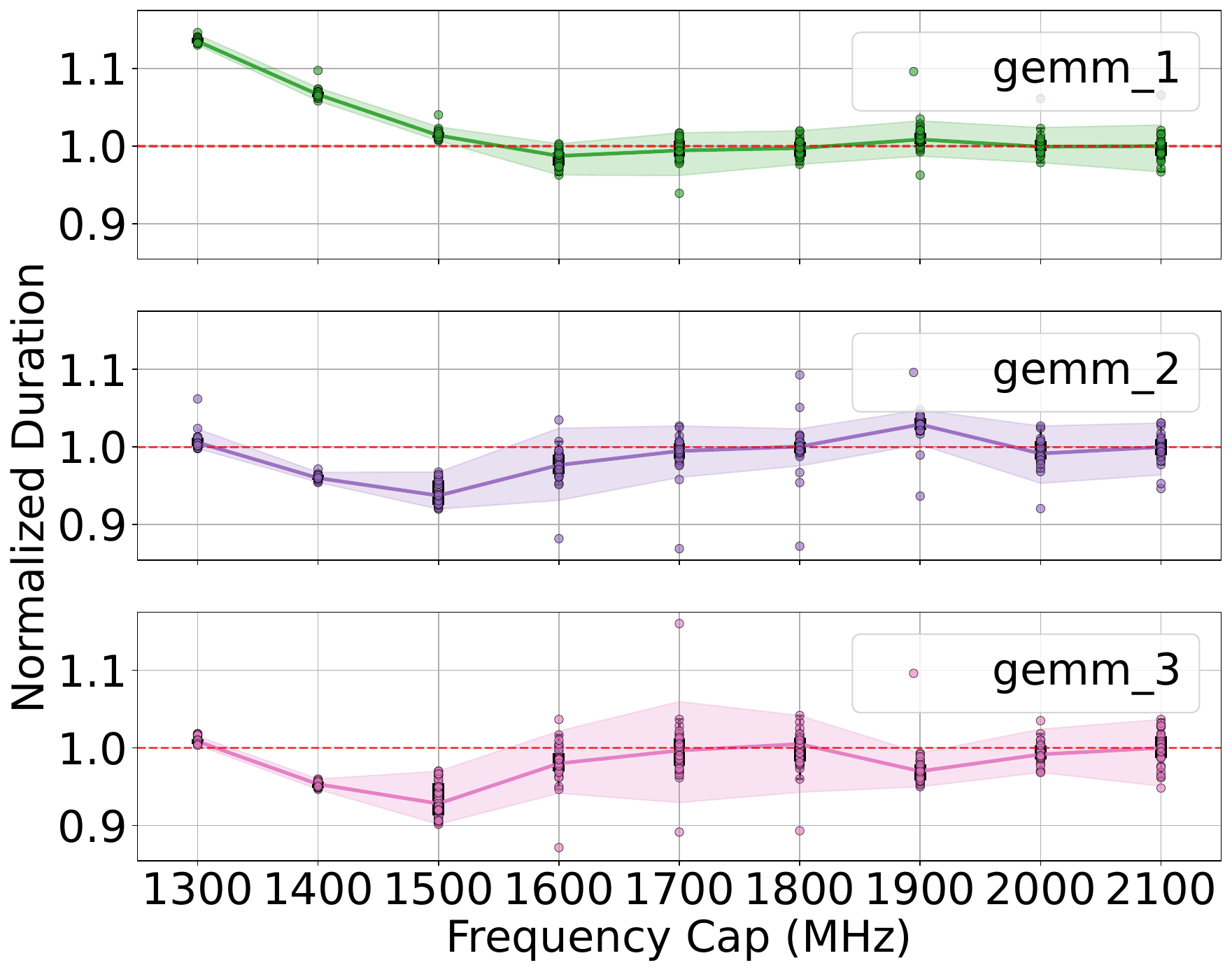}
        \caption{LLaMA3-70B bs1}
    \label{fig:figure-c}
    \end{subfigure}
  \caption{Kernel duration scaling with frequency caps for the top 3 kernels (by kernel duration), normalized to uncapped 2100 MHz runs, for (a) LLaMA3-8B, (b) Mixtral-8x7B, and (c) LLaMA3-70B runs. The kernel distributions show durations for 96 kernel instances across 32 transformer layers and 3 trials for each model.}
  \label{fig:kernel-stability}
  \Description{Kernel duration scaling with frequency caps for the top 3 kernels (by kernel duration) of (a) LLaMA3-8B, (b) Mixtral-8x7B, and (c) LLaMA3-70B runs. The kernel distributions show durations for 96 kernel instances across 32 transformer layers and 3 trials for each model. Durations are normalized to uncapped 2100 MHz runs.}
\end{figure*}

Limiting the maximum frequency reduces the range over which DVFS has to operate, enabling it to react faster\Edit{~\cite{velicka2025methodologygpufrequencyswitching}}.
However, the biggest benefit is that it reduces the extreme \edp{} events that require a substantial corrective solution (i.e., the ones that have the highest amplitude, and shortest duration requirements).
By limiting the maximum frequency, we reduce the occurrence of events where the current operating frequency is considerably higher than the tolerable frequency for the kernel.
Figure~\ref{fig:delta-power-capping} shows the occurrence of \edp{} events with and without frequency limits for a subset of applications in Figure~\ref{fig:oracle-barplots}. 
\Highlight{The results show that with frequency limits, the system continues to utilize lower power overshoots that facilitate computational sprinting but avoid the more significant events that require detrimental corrective action. 
For example, in \mixtralmodel{} with 4096 input tokens and a batch size of 1, the number of \edp{} events increases, but they are of lower magnitude and eliminating events over 1.3$\times$TDP improves performance by 4\%.}
Given the variable nature of these applications, capping the frequency reduces the maximum C$_{AC}$ seen during the run, reducing the delta between the maximum and minimum C$_{AC}$ in the workload and power variability across the workload.

Setting graphics frequency caps also improves system stability, generating more deterministic results.
Figure~\ref{fig:kernel-stability} shows the variability results for the three longest duration kernels, for a subset of the benchmarks in Figure~\ref{fig:oracle-barplots}, with different frequency limits.
The dashed red line at 1.0 shows the mean across all kernel instances across three trials (32 transformer layers per-trial) for a system with no frequency limits (i.e., it can attain a maximum frequency of 2100MHz).
The solid line shows the median performance at that frequency cap relative to the performance of an uncapped system. 
Values below 1.0 indicate a reduction in runtime; values above 1.0 indicate an increase in runtime.

Overall, Figure~\ref{fig:kernel-stability} shows lower graphics frequency limits marginally improve performance and provide more stable kernel operation.
These complex GPUs, running modern ML applications, push the system's physical limits, causing specification-imposed power throttling (Section~\ref{sec:background}).
When and how the system is throttled is a complex algorithm that depends on intrinsic (process variation, power delivery network design, cooling solutions) and extrinsic (ambient temperature, application behavior) factors.
Regardless, pushing the system's physical limits increases performance variability.
Given the frequency of synchronization across applications parallelized across many devices, this variability within and across GPUs could increase overall runtime~\cite{power-stabilization, JainTran2024-pal, Scogland2015-pwrPerspectives, sinha2022notall}. 

\takeawaybox{\begin{itemize}
    \item Setting frequency limits reduces high-magnitude \edp{} events, improving performance by 3-7\% over baseline GPU power management, and is within 3\% of an oracle per-kernel capping. 
    \item Setting frequency limits also provides stable, deterministic operations, reducing transient kernel variability. 
\end{itemize}
}
\subsubsection{Multi-GPU Online Serving}
\label{subsec:online-serving}
As outlined in Section~\ref{sec:design-fw}, we can use the data from offline frequency profiling of models to create a lookup table for optimal $f_{cap}$s for each model and input length. 
When serving requests dynamically, we can use this lookup table to apply $f_{cap}$ corresponding to the effective input length seen every scheduler epoch. 
%
%
%
%
\begin{figure}[t!]
  \centering
  \begin{subfigure}[b]{0.32\textwidth}
    \centering
    \begin{tikzpicture}
      \begin{axis}[kernel stack panel,
        ylabel={Share of prefill kernel time (\%)},
        symbolic x coords={lt2k,2k-4k,4k-8k,8k-16k,16k-32k,32k-64k},
        xticklabels={{$<$2k},{(2k, 4k)},{(4k, 8k)},{(8k, 16k)},{(16k, 32k)},{(32k, 64k)}},
        legend columns=2,
        legend entries={{gemm},{rccl},{fwd\_attn},{layernorm},{scaled\_fp8\_quant},
                        {act\_and\_mul},{rotary},{reshape},{paged\_attn\_reduce},
                        {paged\_attn},{other}},
        ]
        \pgfplotsinvokeforeach{gemm,rccl,fwd_attn,layernorm,scaled_fp8_quant,%
          act_and_mul,rotary,reshape,paged_attn_reduce,paged_attn,other}{%
          \addplot[kb/#1] table[col sep=comma, x=bucket, y=#1]
            {data/kernel_breakdown_llama.csv};
        }
      \end{axis}
    \end{tikzpicture}
    \caption{LLaMA-3.1-405B}
    \label{fig:kernel-breakdown-llama}
  \end{subfigure}
  \hfill
  \begin{subfigure}[b]{0.32\textwidth}
    \centering
    \begin{tikzpicture}
      \begin{axis}[kernel stack panel,
        symbolic x coords={lt2k,2k-4k,4k-8k,8k-16k,16k-32k,32k-64k},
        xticklabels={{$<$2k},{(2k, 4k)},{(4k, 8k)},{(8k, 16k)},{(16k, 32k)},{(32k, 64k)}},
        legend columns=2,
        legend entries={{gemm},{rccl},{mqa\_attn},{gather},
                        {elementwise},{act\_and\_mul},{moe\_sort},{other}},
        ]
        \pgfplotsinvokeforeach{gemm,rccl,mqa_attn,gather,elementwise,%
          act_and_mul,moe_sort,other}{%
          \addplot[kb/#1] table[col sep=comma, x=bucket, y=#1]
            {data/kernel_breakdown_deepseek.csv};
        }
      \end{axis}
    \end{tikzpicture}
    \caption{DeepSeek v3.2}
    \label{fig:kernel-breakdown-deepseek}
  \end{subfigure}
  \hfill
  \begin{subfigure}[b]{0.32\textwidth}
    \centering
    \begin{tikzpicture}
      \begin{axis}[kernel stack panel,
        symbolic x coords={lt1k,1k-2k,2k-4k,4k-8k,8k-16k,16k-32k},
        xticklabels={{$<$1k},{(1k, 2k)},{(2k, 4k)},{(4k, 8k)},{(8k, 16k)},{(16k, 32k)}},
        legend columns=2,
        legend entries={{gemm},{rccl},{fwd\_attn},{layernorm},{scaled\_fp8\_quant},
                        {act\_and\_mul},{rotary},{reshape},{paged\_attn\_reduce},
                        {paged\_attn},{other}},
        ]
        \pgfplotsinvokeforeach{gemm,rccl,fwd_attn,layernorm,scaled_fp8_quant,%
          act_and_mul,rotary,reshape,paged_attn_reduce,paged_attn,other}{%
          \addplot[kb/#1] table[col sep=comma, x=bucket, y=#1]
            {data/kernel_breakdown_mixtral.csv};
        }
      \end{axis}
    \end{tikzpicture}
    \caption{Mistral-Medium-3.5-128B}
    \label{fig:kernel-breakdown-mixtral}
  \end{subfigure}
  \vspace{-1ex}
  \caption{Prefill kernel composition by input-length bucket, normalized to the total prefill kernel time.}
  \Description{Prefill kernel composition by input-length bucket. Each bar is
  normalized to the total prefill kernel time.}
  \label{fig:kernel-breakdown}
  \vspace{-2ex}
\end{figure}
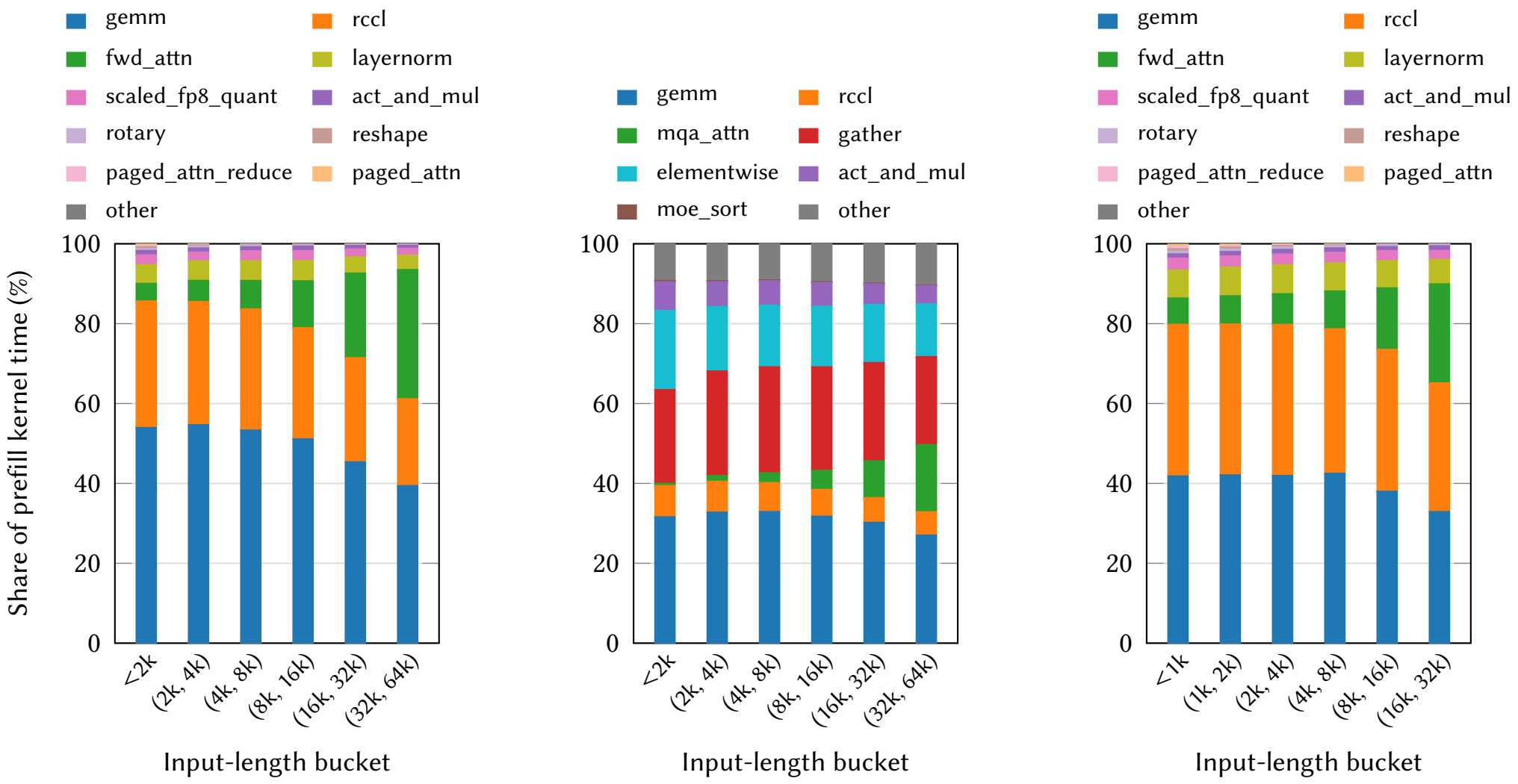

\makeatletter
\def\ftype@table{1}%
\makeatother
%
%
\begin{scriptsize}
\begin{table}[t!]
    \centering
    \caption{Stage 1 output: predicted $f_{cap}$ (MHz) per batched-token
    bucket $b$, and per-cap measured TTFT improvement. We profile DeepSeek v3.2 up to 8k batched tokens due to memory
    capacity issues (OOM).}
    \label{table:fcap-predictions}
    \vspace{-1ex}
    \begin{tabular}{@{}lccccccccc@{}}
        \toprule
        \multirow{2}{*}{\textbf{Bucket $b$}}
          & \multicolumn{3}{c}{\textbf{LLaMA3.1-405B}}
          & \multicolumn{3}{c}{\textbf{DeepSeek v3.2}}
          & \multicolumn{3}{c}{\textbf{Mistral-Medium-3.5-128B}} \\
        \cmidrule(lr){2-4} \cmidrule(lr){5-7} \cmidrule(lr){8-10}
                   & $f_{cap}$ & \%TTFT & mclk & $f_{cap}$ & \%TTFT & mclk & $f_{cap}$ & \%TTFT & mclk \\
        \midrule
        $<$2k      & 1900 & 7.5 & 1550 & 2100 & 0.1 & 1800 & 2000 & 1.0 & default \\
        (2k, 4k)   & 1500 & 9.5 & 1550 & 2000 & 0.0 & 1800 & 1900 & 0.8 & default \\
        (4k, 8k)   & 1600 & 7.5 & 1550 & 1900 & 0.4 & 1800 & 1700 & 3.7 & default \\
        (8k, 16k)  & 1400 & 7.9 & 1550 & \multicolumn{3}{c}{\multirow{3}{*}{OOM}} & 1600 & 3.1 & default \\
        (16k, 32k) & 1300 & 6.7 & 1550 & \multicolumn{3}{c}{}                     & 1400 & 4.4 & default \\
        (32k, 64k) & 1300 & 1.5 & 1550 & \multicolumn{3}{c}{}                     & 1400 & 5.9 & 1550 \\
        \bottomrule
    \end{tabular}
    \Description{Predicted $f_{cap}$ in MHz and the corresponding TTFT improvement for each batched tokens bucket $b$ for LLaMA3.1-405B, DeepSeek v3.2, and Mistral-Medium-3.5-128B.}
    \vspace{-3ex}
\end{table}
\end{scriptsize}

\makeatletter
\def\ftype@table{2}%
\makeatother

\noindent\textbf{Stage 1: Offline frequency profiling.} Figure~\ref{fig:kernel-breakdown} shows the kernel composition for online serving models specified in Table~\ref{tab:inf-workloads}, namely LLaMA3.1-405B, DeepSeek v3.2, and Mistral-Medium-3.5-128B  with varying number of input batched tokens. 
GEMMs dominate overall transformer layer runtime for the dense models (Figures~\ref{fig:kernel-breakdown-llama} and~\ref{fig:kernel-breakdown-mixtral}) across input lengths, with RCCL AllReduce (\texttt{rccl}) and attention (\texttt{fwd\_attn}) being the next largest contributors.
As input length increases for the dense models, attention kernels grow as a percentage of the total runtime, but GEMMs still dominate overall 
as opposed to the dense models where 11 kernels make up a transformer layer, the MoE-based DeepSeek v3.2 model has 61 kernels per layer, with a much more fractured kernel distribution(Figure~\ref{fig:kernel-breakdown-deepseek}).
While GEMMs still make up part of the top-$K$ kernels by runtime for DeepSeek v3.2, the distribution also includes less compute-intensive (lower \CAC{}) kernels like gather, and elementwise kernels. 
As a result, the dense models benefit more from frequency capping since their top $K$ kernels have higher sensitivity to frequency changes. 

Based on the p90 frequencies of the top-$K$ kernels for these models, Table~\ref{table:fcap-predictions} shows the predicted $f_{cap}$ for each batched tokens bucket $b$ for LLaMA3.1-405B, DeepSeek v3.2, and Mistral-Medium-3.5-128B.
For the dense models, the predicted cap decreases linearly as prompts get longer, from 1900\,MHz to 1300\,MHz for LLaMA3.1-405B and from 2000\,MHz to 1400\,MHz for Mistral-Medium-3.5-128B.
This is because larger bucket sizes produce larger running GEMMs with higher \edp{}, which makes the DPM throttle the graphics clock to lower frequencies.
Both dense models also prefer the DPM to keep memory clock states to the default, middle value (1550 MHz).
Conversely, DeepSeek v3.2 prefers both higher frequencies across buckets and a higher memory clock state.
As shown in Figure~\ref{fig:kernel-breakdown-deepseek}, the MoE model's top-$K$ set of kernels includes memory-bound \texttt{gather} and \texttt{elementwise} kernels that do not saturate power. 
Therefore, their p90 frequency stays high, and the maximum frequency selected by Eq~\ref{eq:fcap} is higher than that for dense models.
Moreover, DeepSeek v3.2 also prefers the highest memory clock state since more of its kernels are memory-intensive.

Table~\ref{table:fcap-predictions} also shows the percentage improvement in TTFT for each of these bucket lengths when the corresponding optimal frequency cap is applied.
TTFT improvements show a bell curve: benefits initially increase as prompt length increases, but reduce for very large prompts.  
For a given model, as input length increases, kernels grow in computational demand to cause higher \edp{} and consequently more throttling. 
Frequency capping alleviates the adverse performance impact of this throttling by limiting \edp{}, improving TTFT over baseline as input length increases. 
However, as input lengths grow, the durations of these compute-intensive GEMM kernels also increases, allowing the PM firmware to mitigate \edp{} in the baseline too. 
This means that for larger bucket lengths, such as $>$16k tokens for LLaMA3.1-405B, frequency capping has diminishing returns when compared to the baseline. 
The point at which we get diminishing returns depends on kernel durations and composition of each model.
While these improvements are not necessary as inputs for Stage 2's online serving algorithm, they help explain the results we obtain in Stage 2. 
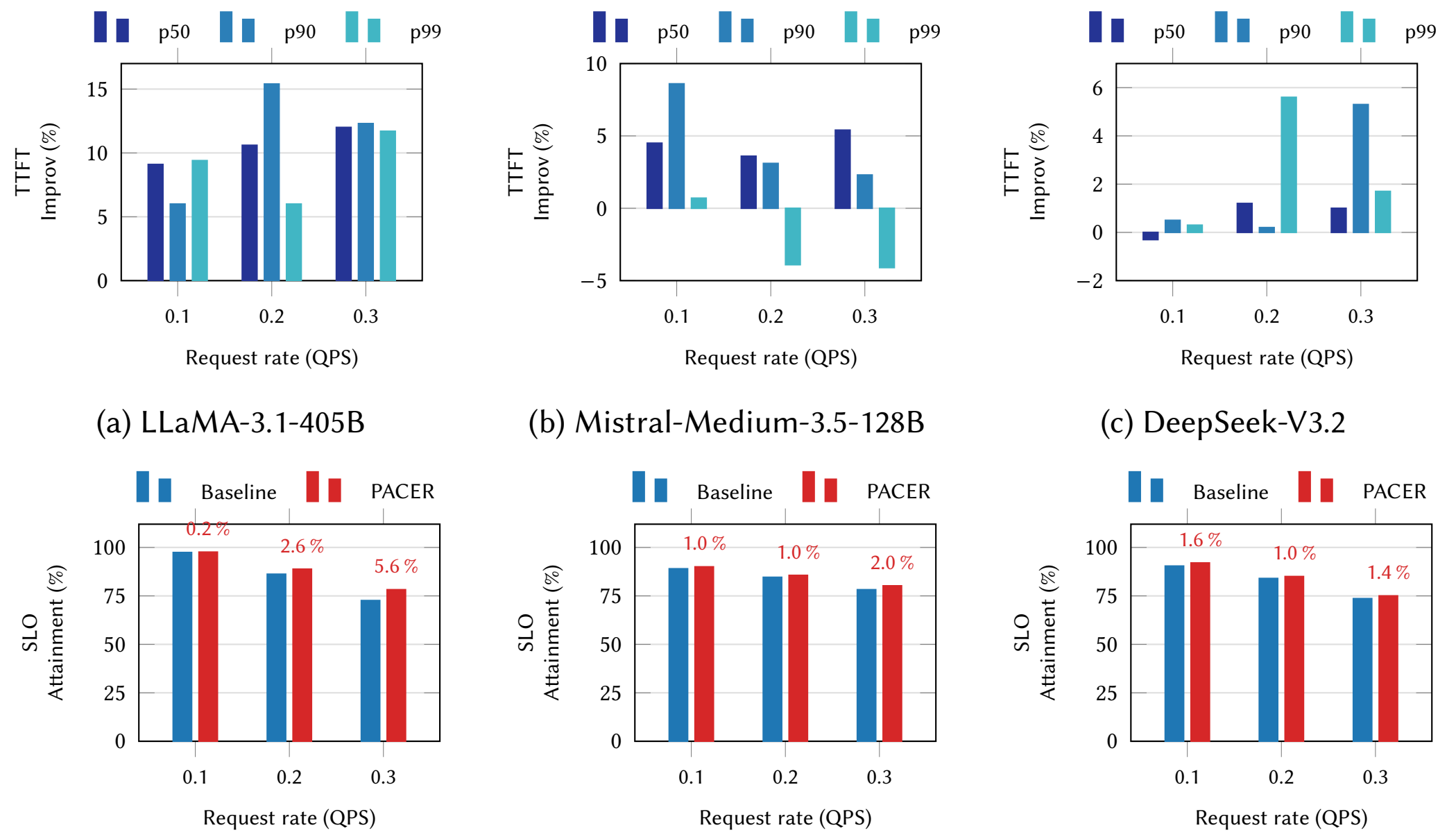
\begin{figure}[t!]
  \centering
  \pgfplotsset{qps bar panel/.append style={
      height=0.82\linewidth,
      tick label style={font=\sffamily\tiny},
      label style={font=\sffamily\tiny},
      ylabel style={font=\sffamily\tiny, align=center},
      legend style={font=\sffamily\tiny},
    }}

  \begin{subfigure}[t]{0.32\textwidth}
    \centering
    \begin{tikzpicture}[font=\sffamily\scriptsize]
      \begin{axis}[qps bar panel,
        ylabel={TTFT\\Improv (\%)},
        bar width=4pt, enlarge x limits=0.3,
        ymin=0, ymax=17, ytick={0,5,10,15},
        legend columns=3,
        ]
        \addplot[draw=ttftpfifty, fill=ttftpfifty]
          table[col sep=comma, x=qps, y=p50] {data/ttft_improvement_llama.csv};
        \addlegendentry{p50}
        \addplot[draw=ttftpninety, fill=ttftpninety]
          table[col sep=comma, x=qps, y=p90] {data/ttft_improvement_llama.csv};
        \addlegendentry{p90}
        \addplot[draw=ttftpninetynine, fill=ttftpninetynine]
          table[col sep=comma, x=qps, y=p99] {data/ttft_improvement_llama.csv};
        \addlegendentry{p99}
      \end{axis}
    \end{tikzpicture}
    \caption{LLaMA-3.1-405B}
    \label{fig:ttft-improvement-llama}
  \end{subfigure}
  \hfill
  \begin{subfigure}[t]{0.32\textwidth}
    \centering
    \begin{tikzpicture}[font=\sffamily\scriptsize]
      \begin{axis}[qps bar panel,
        ylabel={TTFT\\Improv (\%)},
        bar width=4pt, enlarge x limits=0.3,
        ymin=-5, ymax=10, ytick={-5,0,5,10},
        legend columns=3,
        ]
        \addplot[draw=ttftpfifty, fill=ttftpfifty]
          table[col sep=comma, x=qps, y=p50] {data/ttft_improvement_mistral.csv};
        \addlegendentry{p50}
        \addplot[draw=ttftpninety, fill=ttftpninety]
          table[col sep=comma, x=qps, y=p90] {data/ttft_improvement_mistral.csv};
        \addlegendentry{p90}
        \addplot[draw=ttftpninetynine, fill=ttftpninetynine]
          table[col sep=comma, x=qps, y=p99] {data/ttft_improvement_mistral.csv};
        \addlegendentry{p99}
      \end{axis}
    \end{tikzpicture}
    \caption{Mistral-Medium-3.5-128B}
    \label{fig:ttft-improvement-mistral}
  \end{subfigure}
  \hfill
  \begin{subfigure}[t]{0.32\textwidth}
    \centering
    \begin{tikzpicture}[font=\sffamily\scriptsize]
      \begin{axis}[qps bar panel,
        ylabel={TTFT\\Improv (\%)},
        bar width=4pt, enlarge x limits=0.3,
        ymin=-2, ymax=7, ytick={-2,0,2,4,6},
        legend columns=3,
        ]
        \addplot[draw=ttftpfifty, fill=ttftpfifty]
          table[col sep=comma, x=qps, y=p50] {data/ttft_improvement_deepseek.csv};
        \addlegendentry{p50}
        \addplot[draw=ttftpninety, fill=ttftpninety]
          table[col sep=comma, x=qps, y=p90] {data/ttft_improvement_deepseek.csv};
        \addlegendentry{p90}
        \addplot[draw=ttftpninetynine, fill=ttftpninetynine]
          table[col sep=comma, x=qps, y=p99] {data/ttft_improvement_deepseek.csv};
        \addlegendentry{p99}
      \end{axis}
    \end{tikzpicture}
    \caption{DeepSeek-V3.2}
    \label{fig:ttft-improvement-deepseek}
  \end{subfigure}

  \vspace{1ex}

  \begin{subfigure}[t]{0.32\textwidth}
    \centering
    \begin{tikzpicture}[font=\sffamily\scriptsize]
      \begin{axis}[qps bar panel,
        ylabel={SLO\\Attainment (\%)},
        bar width=5pt, enlarge x limits=0.3,
        ymin=0, ymax=112, ytick={0,25,50,75,100},
        legend columns=2,
        nodes near coords align={vertical},
        ]
        \addplot[draw=baselinecol, fill=baselinecol]
          table[col sep=comma, x=qps, y=baseline] {data/slo_attainment_llama.csv};
        \addlegendentry{Baseline}
        \addplot[draw=solutioncol, fill=solutioncol,
                 point meta=explicit symbolic,
                 nodes near coords={\pgfplotspointmeta\,\%},
                 every node near coord/.append style={
                   font=\sffamily\tiny, color=solutioncol, anchor=south,
                   yshift=1pt}]
          table[col sep=comma, x=qps, y=solution, meta=delta]
          {data/slo_attainment_llama.csv};
        \addlegendentry{\SolutionName{}}
      \end{axis}
    \end{tikzpicture}
    \caption{LLaMA-3.1-405B}
    \label{fig:slo-attainment-llama}
  \end{subfigure}
  \hfill
  \begin{subfigure}[t]{0.32\textwidth}
    \centering
    \begin{tikzpicture}[font=\sffamily\scriptsize]
      \begin{axis}[qps bar panel,
        ylabel={SLO\\Attainment (\%)},
        bar width=5pt, enlarge x limits=0.3,
        ymin=0, ymax=112, ytick={0,25,50,75,100},
        legend columns=2,
        nodes near coords align={vertical},
        ]
        \addplot[draw=baselinecol, fill=baselinecol]
          table[col sep=comma, x=qps, y=baseline] {data/slo_attainment_mistral.csv};
        \addlegendentry{Baseline}
        \addplot[draw=solutioncol, fill=solutioncol,
                 point meta=explicit symbolic,
                 nodes near coords={\pgfplotspointmeta\,\%},
                 every node near coord/.append style={
                   font=\sffamily\tiny, color=solutioncol, anchor=south,
                   yshift=1pt}]
          table[col sep=comma, x=qps, y=solution, meta=delta]
          {data/slo_attainment_mistral.csv};
        \addlegendentry{\SolutionName{}}
      \end{axis}
    \end{tikzpicture}
    \caption{Mistral-Medium-3.5-128B}
    \label{fig:slo-attainment-mistral}
  \end{subfigure}
  \hfill
  \begin{subfigure}[t]{0.32\textwidth}
    \centering
    \begin{tikzpicture}[font=\sffamily\scriptsize]
      \begin{axis}[qps bar panel,
        ylabel={SLO\\Attainment (\%)},
        bar width=5pt, enlarge x limits=0.3,
        ymin=0, ymax=112, ytick={0,25,50,75,100},
        legend columns=2,
        nodes near coords align={vertical},
        ]
        \addplot[draw=baselinecol, fill=baselinecol]
          table[col sep=comma, x=qps, y=baseline] {data/slo_attainment_deepseek.csv};
        \addlegendentry{Baseline}
        \addplot[draw=solutioncol, fill=solutioncol,
                 point meta=explicit symbolic,
                 nodes near coords={\pgfplotspointmeta\,\%},
                 every node near coord/.append style={
                   font=\sffamily\tiny, color=solutioncol, anchor=south,
                   yshift=1pt}]
          table[col sep=comma, x=qps, y=solution, meta=delta]
          {data/slo_attainment_deepseek.csv};
        \addlegendentry{\SolutionName{}}
      \end{axis}
    \end{tikzpicture}
    \caption{DeepSeek-V3.2}
    \label{fig:slo-attainment-deepseek}
  \end{subfigure}

  \caption{Online serving results. Top row: reduction in TTFT with
  \SolutionName{} over baseline at each percentile (p50/p90/p99); higher is better.
  Bottom row: fraction of requests meeting the TTFT SLO, annotated
  with \SolutionName{}'s percentage gain over baseline.}
  \Description{Online serving results. Top row: reduction in TTFT with
  \SolutionName{} over baseline at each percentile (p50/p90/p99); higher is better.
  Bottom row: fraction of requests meeting the TTFT SLO, annotated with
  \SolutionName{}'s gain over baseline in percentage points.}
  \label{fig:ttft-slo}
  \vspace{-4ex}
\end{figure}

\noindent\textbf{Stage 2: Online $f_{cap}$ deployment.} 
Using the $f_{cap}$ model table in Stage 1, we then deploy these frequency caps dynamically, as detailed in Section~\ref{sec:design-fw}. 
Figure~\ref{fig:ttft-slo} shows online serving results at varying request rates, comparing \SolutionName{} against baseline behavior, where all frequencies are uncapped. 
For LLaMA3.1-405B (Figures~\ref{fig:ttft-improvement-llama} and~\ref{fig:slo-attainment-llama}), \SolutionName{} reduces median TTFT by 9--12\% across request rates.
Frequency capping has unequal benefits depending on prompt length, so the dynamic batched token lengths at different request rates can affect how much TTFT improves. 
However, we observed that the p50 batched tokens per scheduler step is nearly rate-independent -- the median token length per step grows from 8.5k to
10k tokens as QPS is increased from 0.1 to 0.3. 
Therefore, we only see a marginal improvement in p50 TTFT as request rate is increased. 
SLO attainment improves from 97.4\% to 97.6\% at 0.1 QPS and from 72.6\% to 78.2\% at 0.3 QPS.
Although the per-request TTFT reduction is roughly constant with load, its effect on attainment grows with QPS. 
At 0.1 QPS, the baseline already meets the SLO target for 97.4\% of requests and there is little headroom left to recover, whereas at 0.3 QPS, queueing pushes over a quarter of requests past the threshold and \SolutionName{} improves attainment by 6\%. 

Mistral-Medium-3.5-128B shows the same effect as LLaMA3.1-405B, improving median TTFT by 4--5\% and SLO attainment by up to 2\% (Figures~\ref{fig:ttft-improvement-mistral} and~\ref{fig:slo-attainment-mistral}).
The lower percentages of improvement track with the offline inference results in Table~\ref{table:fcap-predictions}, which showed around 3--6\% improvements for different buckets for the Mistral model. 
The tail (p99) of TTFT is iso-performance to the baseline or degrades with \SolutionName{} because large requests operate in the diminishing returns region of the bell curve, and queueing delays dominate TTFT more than frequency changes. 

DeepSeek v3.2 benefits least, improving median TTFT by at most 1.2\% and SLO attainment by about 2\% (Figures~\ref{fig:ttft-improvement-deepseek} and Figures~\ref{fig:slo-attainment-deepseek}). 
This follows directly from the $f_{cap}$ model and Stage 1's kernel composition. 
DeepSeek's predicted caps sit at or near the uncapped 2100\,MHz ceiling for the buckets it can serve across 8 GPUs, so \SolutionName{} leaves the firmware close to its default operating state. 
With 61 kernels per layer and a large fraction of time running low-\CAC{} kernels, DeepSeek v3.2 does not suffer from as much \edp{} events that would warrant aggressive frequency capping or \edp{} management. 
MoE models with fractured kernel distributions are therefore poor candidates for frequency capping, and \SolutionName{} accordingly retains performance close to the baseline behavior. 

\takeawaybox{\begin{itemize}
    \item For dense models, \SolutionName{} improves TTFT by up to 12\% and SLO attainment by up to 6\%. 
    \item Each model's benefits track its distribution of frequency-sensitive kernels: GEMM-dominated dense models gain the most, while MoE models with fractured kernel distributions see much smaller improvements.
\end{itemize}
}

\section{Related Work}
\label{sec:previous}
Other works have also noted the existence of \edp{}.
Patel, et al. noted GPU \edp{} complicate power oversubscription solutions, which rely on fixed power budgets to monitor and manage power oversubscription at the rack level or higher~\cite{Patel-CAL}.
Later work from Patel and others shows power peaks interfere with power caps~\cite{LLM-ASPLOS, dynamollm}, resulting in noticeable overshoots beyond the power cap.
Notably, the work by Choukse, et al.~\cite{power-stabilization} set a power floor of 0.9$\times$TDP, their dynamic power range is 20\% given a minimum EDP point of 1.1$\times$TDP.
Existing work identifies and qualitatively discusses the ramifications of power overshoots~\cite{power-stabilization, LLM-ASPLOS, Patel-CAL, dynamollm,  Wallossek2025-autoDownReg}.
Conversely, our work performs a detailed \Highlight{quantification} on the benefits and costs of peak power management on \Edit{modern GPU workloads}.

Some works have identified frequency capping as a method to control power consumption of workloads, thereby reducing workload energy, with tolerable loss in performance. 
Kakolyris et al.~\cite{SLOAwareDVFS} designed an offline profiling based frequency selection algorithm for inference serving on NVIDIA GPUs, but their objective was to reduce energy consumption while meeting TBT SLOs in decode, not to reduce variability or improve performance or SLO attainment during prefill. 
Thus, while they use frequency limits as a control parameter similar to \SolutionName{}, our approach is complementary to theirs and identifies potential for performance \emph{improvements} and transient jitter reduction by managing the same control variable, frequency limits, in prefill while they manage it in the decode phase.  
Other work~\cite{Guerreiro-appClasses, MI250XFreqCap} also look at dynamic frequency selection, but for scientific workloads or smaller micro-benchmarks, not LLMs which we have impose unique constraints on the system's power management (Section~\ref{subsec:exbehavior-ml}). 
Jain et al.~\cite{Minos-sigmetrics} do quantify \edp{} across workload types, but they use the power distribution of workloads to group them into classes, so known frequency settings from previously profiled workloads can be applied to a new, unseen workload. 
Moreover, they assume a single frequency cap works for a workload, while we show that frequencies are model- and input-dependent, with kernel distributions and their \CAC{} affecting \edp{} and consequently the optimal frequencies for the workload.
Instead, \SolutionName{} proposes a one-time baseline profiling of kernel composition and frequencies to identify optimal frequencies for each model and input during online serving.
Hyperscalers dedicate entire clusters to serving the same model, so we believe that \SolutionName{}'s offline profiling step is a one-time cost. 

Other work uses intra-device GPU parallelism to increase existing hardware usage~\cite{AutoscalingGranularity, PodAttention, NanoFlow}.
Although these works improve performance by more effectively sharing resources, they do not consider power management or \edp{}.
Thus, they focus on different problems.
\vspace{-1ex}
\section{Conclusion}
\label{sec:conclusion}

Computing systems are being driven by the voracious scaling needs of workloads, especially ML.
Unfortunately, these workloads frequently cause \edp{}, leading to jittery power and frequency behavior that can impact system stability. 
To understand the impact of \edp{}, we characterized it across graph analytics, HPC, and ML workloads.
Our results show that fine-grained variations in compute requirements, especially for ML workloads, make stable power management extremely difficult.
While vendors have added complex DPM support to prevent thermal meltdown and robust VRs for power delivery, running small intra- and inter-kernel phases with different power behavior in rapid succession still leads to \edp{}, causing power and frequency variability, often degrading performance for these workloads.
To manage these \edp{} and power variations, we examined a fine-grained hardware mechanism to strictly clamp power within limits and a firmware frequency capping solution, \SolutionName{}, that limits maximum operating frequency limits.
Overall, across 
popular ML workloads, our approaches improve performance by 3-12\% on state-of-the-art, production-grade systems and are within 3\% of an oracular solution.
More generally, our work highlights how power 
must be intelligently managed across 
peak power, average power, and \edp{} 
to meet key performance goals across diverse workloads. 
\printbibliography



\end{document}